\documentclass[prd,showpacs,preprintnumbers,amsmath,amssymb]{revtex4}

\usepackage{graphics}
\usepackage{epsfig}
\usepackage{subfigure}
\usepackage{dcolumn}
\usepackage{bm}
\usepackage{color}

\begin{document}
\draft
\title{{\bf Are Y(4260) and {\rm Z$_2^{+}$(4250)} ${\rm D_1D}$ or ${\rm D_0D^{*}}$ Hadronic Molecules?}}

\author{Gui-Jun Ding}

\affiliation{\centerline{Department of Modern
Physics,}\centerline{University of Science and Technology of
China,Hefei, Anhui 230026, China}}


\begin{abstract}
\vskip0.5cm

 In this work, we have investigated whether Y(4260) and
${\rm Z^{+}_2(4250)}$ could be ${\rm D_1D}$ or ${\rm D_0D^{*}}$
molecules in the framework of meson exchange model. The off-diagonal
interaction induced by $\pi$ exchange plays a dominant role. The
$\sigma$ exchange has been taken into account, which leads to
diagonal interaction. The contribution of $\sigma$ exchange is not
favorable to the formation of molecular state ${\rm with~
I^{G}(J^{PC})=0^{-}(1^{--})}$, however, it is beneficial to the
binding of molecule ${\rm with~ I^{G}(J^{P})=1^{-}(1^{-})}$. Light
vector meson exchange leads to diagonal interaction as well. For
${\rm Z^{+}_2(4250)}$, the contribution from $\rho$ and $\omega$
exchange almost cancels each other. For the currently allowed values
of the effective coupling constants and a reasonable cutoff
$\Lambda$ in the range 1-2 GeV, We find that Y(4260) could be
accommodated as a ${\rm D_1D}$ and ${\rm D_0D^{*}}$ molecule,
whereas the interpretation of ${\rm Z^{+}_2(4250)}$ as a ${\rm
D_1D}$ or ${\rm D_0D^{*}}$ molecule is disfavored. The bottom analog
of Y(4260) and ${\rm Z^{+}_2(4250)}$ may exist, and the most
promising channels to discovery them are $\pi^{+}\pi^{-}\Upsilon$
and $\pi^{+}\chi_{b1}$ respectively.

\end{abstract}

\maketitle

\section{introduction}
In the past years, a number of charmonium-like  X, Y, Z states have
been observed, which stimulate a lot of discussion about the
structures and properties of these resonances. In particular, the
Z$^{+}$(4430) observed in the $\pi^{+}\psi'$ invariant
spectrum\cite{:2007wga} carries one unit electric charge.
Consequently, it can not be simple charmonium.  Recently, two new
resonance-like structures Z$_1^{+}$(4051) and Z$_2^{+}$(4250) in the
$\pi^{+}\chi_{c1}$ mass distribution in exclusive
$\overline{B}^{0}\rightarrow K^{-}\pi^{+}\chi_{c1}$ have been
reported by the Belle collaboration\cite{Mizuk:2008me}. Their masses
and widths are determined to be ${\rm
M_1=(4051\pm14^{+20}_{-41})MeV}$, ${\rm
\Gamma_1=(82^{+21+47}_{-17-22})MeV}$, ${\rm
M_2=(4248^{+44+180}_{-29-35})MeV}$ and ${\rm
\Gamma_2=(177^{+54+316}_{-39-61})MeV}$ respectively, with the
product branching fractions ${\cal B}(\overline{B}^{0}\rightarrow
K^{-}Z^{+}_{1,2})\times{\cal
B}(Z^{+}_{1,2}\rightarrow\pi^{+}\chi_{c1})=(3.0^{+1.5+3.7}_{-0.8-1.6})\times10^{-5}$
and $(4.0^{+2.3+19.7}_{-0.9-0.5})\times10^{-5}$ respectively. Both
Z$_1^{+}$(4051) and Z$_2^{+}$(4250) carry one unit electric charge
like Z$^{+}$(4430), hence they must be states beyond quark model, if
these states are confirmed in future. Since $\pi^{+}$ is an
isovector with negative G-parity, and $\chi_{c1}$ is a isospin
singlet with positive G-parity, the quantum numbers of
Z$_1^{+}$(4051) and Z$_2^{+}$(4250) are ${\rm I^{G}=1^{-}}$. It is
remarkable that some states are in the vicinity of the S-wave
threshold of two charmed mesons, e.g., X(3872) and Z$^{+}$(4430) are
very close to the thresholds of ${\rm D^{*}D}$ and ${\rm D_1D^{*}}$
respectively, therefore it is tempting to interpret these states as
molecular states\cite{X3872,Z4430}. Particularly, Y(4260) and
Z$_2^{+}$(4250) are close to the ${\rm D_1D}$ and ${\rm D_0D^{*}}$
thresholds, which inspires the theoretical interpretations of
Y(4260) as a ${\rm D_0D^{*}}$ molecule\cite{Albuquerque:2008up} and
Z$_2^{+}$(4250) as a ${\rm D_1D}$ molecule\cite{Lee:2008gn}.

Y(4260) was reported by the Babar collaboration in the
$\pi^{+}\pi^{-}J/\psi$ invariant spectrum of the reaction
$e^{+}e^{-}\rightarrow\gamma_{ISR}\pi^{+}\pi^{-}J/\psi$\cite{Aubert:2005rm},
which has been confirmed by both the CLEO and Belle
collaboration\cite{He:2006kg,:2007sj}. A fit to the peak with a
single Breit-Winger resonance shape yields a mass ${\rm
M=(4259\pm10)MeV}$ and the full width ${\rm \Gamma=(88\pm24)MeV}$.
Evidently the state is a vector with $c\bar{c}$ flavor, and its
quantum numbers are determined to be ${\rm
I^{G}(J^{PC})=0^-(1^{--})}$. Although it is above the threshold for
decaying into ${\rm D\overline{D}}$, ${\rm
D\overline{D}^{\,*}(D^{*}\overline{D})}$ or ${\rm
D^{*}\overline{D}^{*}}$ meson pairs, there is no evidence for
Y(4260) in these
channels\cite{Abe:2006fj,Pakhlova:2008zz,Aubert:2008pa}. Therefore
Y(4260) appears not to be a canonical charmonium.

The observation of the Y(4260) has sparked many theoretical
speculations. It has variously been identified as a conventional
$\psi(4S)$ based on a relativistic quark
model\cite{LlanesEstrada:2005hz}, a tetraquark $c\bar{c}s\bar{s}$
state\cite{Maiani:2005pe} which decays predominantly into ${\rm
D_s\overline{D}_s}$, or a charmonium hybrid\cite{Y4260}. The data on
$e^{+}e^{-}\rightarrow {\rm D_s\bar{D}_s}$ show a peaking above
threshold around 4 GeV but no evidence of affinity for a structure
at 4.26GeV\cite{Dubrovin:2007gu}. If these data are confirmed, then
the interpretation of Y(4260) as a $c\bar{c}s\bar{s}$ tetraquark
would be ruled out. Moreover, dynamical calculation of tetraquark
states indicated that Y(4260) can not be interpreted as P-wave
$1^{--}$ state of charm-strange diquark-antidiquark, because the
corresponding mass is found to be 200 MeV
heavier\cite{Ebert:2005nc}. Although the charmonium hybrid is a very
attractive interpretation, the Lattice QCD simulations predict that
lightest charmonium hybrid is about 4.4GeV\cite{lat-ch}, which is
very close to the new charmonium-like state Y(4360)\cite{:2007ea}.
As has been proposed in Ref.\cite{Ding:2007rg}, a possible
resolution to this issue is that Y(4360) is the candidate of
charmonium hybrid, while Y(4260) is a ${\rm D_1D}$ hadronic
molecule.

In Ref.\cite{Swanson:2005tq}, Swanson emphasized that we should
examine the ${\rm D_1D}$ molecular interpretation before finally
concluding that Y(4260) is a charmonium hybrid. Furthermore, he
pointed out that $\pi$ exchange does not lead to a diagonal
interaction in the ${\rm D_1D}$ channel, and certain novel mechanism
such as off-diagonal interaction may be required. In
Refs.\cite{Close:2008hv,Close:2007ny}, Close showed that parity
conservation requires the $\pi$ vertex to link ${\rm
D\leftrightarrow D^{*}}$ and ${\rm D_1\leftrightarrow D_0}$, then
the $\pi$ exchange gives an off-diagonal potential linking ${\rm
D_1\overline{D}\leftrightarrow D_0\overline{D}^{*}}$ or ${\rm
D\overline{D}_1\leftrightarrow D^{*}\overline{D}_0}$. This $\pi$
exchange attraction possibly results in a $1^{--}$ hadronic molecule
near the ${\rm D_1D}$ threshold. In this work, we shall investigate
whether Y(4260) and Z$_2^{+}(4250)$ could be hadronic molecule due
to the off-diagonal $\pi$ exchange effect in the framework of heavy
quark effective theory. The contribution of $\sigma$ exchange has
been considered, which results in diagonal interaction. The light
vector mesons $\rho$ and $\omega$ exchange is discussed as well.

The paper is organized as follows. In section II, we present the
formalism to include both heavy meson and anti-meson fields in the
heavy meson chiral perturbation theory(HM$\chi$PT), and the complete
Lagrangian is written out explicitly. Section III illustrates the
systematic procedure for converting a general T-matrix into an
equivalent potential operator. Later we follow this to derive the
effective potential. In section IV, we present both the diagonal and
non-diagonal potential related with Y(4260) and ${\rm
Z^{+}_2(4250)}$.
In section V, we investigate the possible bound states of the ${\rm
D_1D}$ and ${\rm D_0D^{*}}$ system by solving the coupled-channel
Schr$\ddot{\rm o}$dinger equations, and the structures of Y(4260)
and ${\rm Z^{+}_2(4250)}$ are discussed. Moreover, the bottom analog
of Y(4260) and ${\rm Z^{+}_2(4250)}$ is studied. We present our
conclusions and some relevant discussions in Section VI. Finally,
the potential from $\rho$ and $\omega$ exchange is shown in the
Appendix.

\section{Formalism for the system containing both meson and anti-meson fields in HM$\chi$PT}
The strong interaction between pseudo-Goldstone bosons and the
mesons containing a heavy quark is described by the so-called heavy
meson chiral perturbation theory
(HM$\chi$PT)\cite{hmkpt1,hmkpt2,hmkpt3}. The heavy meson chiral
perturbation theory is constructed starting from the spin-flavor
symmetry occurring in QCD in the infinite heavy quark mass limit,
and from the chiral symmetry valid in the massless limit for the
light quarks. In HM$\chi$PT, the heavy-light meson field appears in
a covariant form, which is represented by a $4\times 4$ Dirac-type
matrix. The negative and positive parity doublets containing a heavy
quark $Q$ and a light anti-quark of flavor $a$, can be respectively
described by the superfields $H_a$, $S_a$ and $T^{\mu}_a$ as follows
\begin{eqnarray}
\nonumber&&H^{(Q)}_a=\frac{1+\not
v}{2}\;[P^{*(Q)\mu}_a\gamma_{\mu}-P^{(Q)}_a\gamma_5]\\
\nonumber&&S^{(Q)}_a=\frac{1+\not
v}{2}[P^{'(Q)\mu}_{1a}\gamma_{\mu}\gamma_5-P^{*(Q)}_{0a}]\\
\label{1}&&T^{(Q)\mu}_a=\frac{1+\not
v}{2}[P^{*(Q)\mu\nu}_{2a}\gamma_{\nu}-\sqrt{\frac{3}{2}}P^{(Q)}_{1a\nu}\gamma_5(g^{\mu\nu}-\frac{1}{3}\gamma^{\nu}(\gamma^{\mu}-v^{\mu}))]
\end{eqnarray}

The above various operators annihilate mesons of four-velocity $v$
which is conserved in strong interaction processes. The heavy field
operators contain a factor $\sqrt{\rm M_P}$ and have dimension 3/2.
Under a heavy quark spin $SU(2)$ transformation $S$ and a generic
light flavor transformation $U$ (i.e., $U\in SU(3)$)
\begin{equation}
\label{2}H^{(Q)}_a\rightarrow
SH^{(Q)}_bU^{\dagger}_{ba},~~S^{(Q)}_a\rightarrow
SS^{(Q)}_bU^{\dagger}_{ba},~~T^{(Q)\mu}_a\rightarrow
ST^{(Q)\mu}_bU^{\dagger}_{ba}
\end{equation}
The conjugate field, which creates heavy-light mesons containing a
heavy quark $Q$ and a light anti-quark of flavor $a$, is defined as
\begin{equation}
\label{3} \overline{H}^{\,(Q)}_a=\gamma_0H^{(Q)\dagger}_a\gamma_0,~~
\overline{S}^{\,(Q)}_a=\gamma_0S^{(Q)\dagger}_a\gamma_0,~~\overline{T}^{\,(Q)\mu}_a\equiv\gamma_0T^{(Q)\mu\dagger}_a\gamma_0
\end{equation}
which transforms under $S$ and $U$ as
\begin{equation}
\label{4}\overline{H}^{\,(Q)}_a\rightarrow
U_{ab}\overline{H}^{\,(Q)}_bS^{\dagger},~~
\overline{S}^{\,(Q)}_a\rightarrow
U_{ab}\overline{S}^{\,(Q)}_bS^{\dagger},~~
\overline{T}^{\,(Q)\mu}_a\rightarrow
U_{ab}\overline{T}^{\,(Q)\mu}_bS^{\dagger}
\end{equation}
The octet of light pseudoscalar mesons can be introduced using the
non-linear representation $\Sigma=\xi^2$ and $\xi=\exp(i{\cal
M}/f_{\pi})$ with $f_{\pi}=132$ MeV. The matrix ${\cal M}$ contains
$\pi$, $K$, $\eta$ fields, which is  a $3\times 3$ hermitian and
traceless matrix
\begin{equation}
\label{5}{\cal M}=\left(
\begin{array}{ccc}
\frac{\pi^0}{\sqrt{2}}+\frac{\eta}{\sqrt{6}}&\pi^{+}&K^{+}\\
\pi^{-}&-\frac{\pi^{0}}{\sqrt{2}}+\frac{\eta}{\sqrt{6}}&K^{0}\\
K^{-}&\overline{K}^{0}&-\sqrt{\frac{2}{3}}\,\eta
\end{array}
\right)
\end{equation}
Under the chiral symmetry, the field $\xi$ transforms as
\begin{equation}
\label{6}\xi\rightarrow g_L\xi U^{\dagger}=U\xi g^{\dagger}_R
\end{equation}
where $g_L$ and $g_R$ are left-handed and right-handed global
$SU(3)$ transformation respectively.

The effective QCD Lagrangian is constructed by imposing invariance
under both heavy quark spin-flavor transformation and chiral
transformation, it
is\cite{hmkpt1,hmkpt2,hmkpt3,Casalbuoni:1996pg,Falk:1992cx}
\begin{eqnarray}
\nonumber&&{\cal L}_{\rm P}=ig\langle H^{(Q)}_b{\cal
A}\!\!\!\slash_{ba}\gamma_5\overline{H}^{\,({Q})}_a\rangle+ik\langle
T^{(Q)\mu}_{b}{\cal
A}\!\!\!\slash_{ba}\gamma_5\overline{T}^{\,(Q)}_{a\mu}\rangle+i\tilde{k}\langle
S^{(Q)}_b{\cal
A}\!\!\!\slash_{ba}\gamma_5\overline{S}^{\,(Q)}_a\rangle+\Big[ih\langle
S^{(Q)}_b{\cal
A}\!\!\!\slash_{ba}\gamma_5\overline{H}^{\,(Q)}_a\rangle\\
\label{7}&&+i\tilde{h}\langle T^{(Q)\mu}_b{\cal A}_{\mu
ba}\gamma_5\overline{S}^{\,(Q)}_a\rangle+i\frac{h_1}{\Lambda_{\chi}}\langle
T^{(Q)\mu}_b(D_{\mu}{\cal
A}\!\!\!\slash)_{ba}\gamma_5\overline{H}^{\,(Q)}_a\rangle+i\frac{h_2}{\Lambda_{\chi}}\langle
T^{(Q)\mu}_b(D\!\!\!\!/ {\cal
A}_{\mu})_{ba}\gamma_5\overline{H}^{\,(Q)}_a\rangle+h.c.\Big]
\end{eqnarray}
where $\langle...\rangle$ means trace over the $4\times4$ matrices,
the covariant derivative $D_{\mu}=\partial_{\mu}+{\cal V}_{\mu}$,
the vector current ${\cal V}_{\mu}$ and the axial current ${\cal
A}_{\mu}$ are defined by
\begin{eqnarray}
\nonumber&&{\cal
V}_{\mu}=\frac{1}{2}(\xi^{\dagger}\partial_{\mu}\xi+\xi\partial_{\mu}\xi^{\dagger})\\
\label{8}&&{\cal
A}_{\mu}=\frac{1}{2}(\xi^{\dagger}\partial_{\mu}\xi-\xi\partial_{\mu}\xi^{\dagger})
\end{eqnarray}
In order to describe mesons containing heavy anti-quark
$\overline{Q}$, we have to introduce six new fields
$P^{*(\overline{Q})}_{a\mu}$, $P^{(\overline{Q})}_a$,
$P^{'(\overline{Q})}_{1a\mu}$, $P^{*(\overline{Q})}_{0a}$,
$P^{*(\overline{Q})}_{2a\mu\nu}$ and $P^{(\overline{Q})}_{1a\mu}$
which destroy mesons containing a heavy anti-quark $\overline{Q}$.
The phase of the field $P^{*(\overline{Q})}_{a\mu}$ relative to
$P^{*({Q})}_{a\mu}$, $P^{(\overline{Q})}_a$ to $P^{({Q})}_a$ {etc}
can be fixed by the following charge conjugation convention
\begin{eqnarray}
\nonumber&&P^{*(\overline{Q})}_{a\mu}=-{\cal
C}P^{*({Q})}_{a\mu}{\cal C}^{-1},~~~P^{(\overline{Q})}_a={\cal
C}P^{({Q})}_a{\cal C}^{-1},~~~P^{'(\overline{Q})}_{1a\mu}={\cal
C}P^{'({Q})}_{1a\mu}{\cal C}^{-1},\\
\nonumber&&P^{*(\overline{Q})}_{0a}={\cal C}P^{*({Q})}_{0a}{\cal
C}^{-1},~~~P^{*(\overline{Q})}_{2a\mu\nu}=-{\cal
C}P^{*({Q})}_{2a\mu\nu}{\cal
C}^{-1},~~~P^{(\overline{Q})}_{1a\mu}={\cal C}P^{({Q})}_{1a\mu}{\cal
C}^{-1},\\
\label{9}&&{\cal C}\xi{\cal C}^{-1}=\xi^{T},~~~~~~~~~~~~~~~{\cal
C}{\cal V}_{\mu}{\cal C}^{-1}=-{\cal V}_{\mu}^{T},~~~~~~~~~{\cal
C}{\cal A}_{\mu}{\cal C}^{-1}={\cal A}_{\mu}^{T}
\end{eqnarray}
The mesons containing a heavy anti-quark $\overline{Q}$ and a light
quark of flavor $a$ can be included into the theory by applying the
charge conjugation operation to the above heavy-light meson
superfields $H^{(Q)}_a$, $S^{(Q)}_a$ and
$T^{(Q)}_{a\mu}$\cite{Grinstein:1992qt}
\begin{eqnarray}
\nonumber&&H^{(\overline{Q})}_a=C({\cal C}H^{(Q)}_a{\cal
C}^{-1})^{T}C^{-1}=[P^{*(\overline{Q})\mu}_{a}\gamma_{\mu}-P^{(\overline{Q})}_a\gamma_5]\frac{1-\not
v}{2}\\
\nonumber&&S^{(\overline{Q})}_a=C({\cal C}S^{(Q)}_a{\cal
C}^{-1})^{T}C^{-1}=[P^{'(\overline{Q})\mu}_{1a}\gamma_{\mu}\gamma_5-P^{*(\overline{Q})}_{0a}]\frac{1-\not
v}{2}\\
\label{10}&&T^{(\overline{Q})\mu}_{a}=C({\cal C}T^{(Q)}_{a\mu}{\cal
C}^{-1})^{T}C^{-1}=[P^{(\overline{Q})\mu\nu}_{2a}\gamma_{\nu}-\sqrt{\frac{3}{2}}P^{(\overline{Q})}_{1a\nu}\gamma_5(g^{\mu\nu}-\frac{1}{3}(\gamma^{\mu}-v^{\mu})\gamma^{\nu})]\frac{1-\not
v}{2}
\end{eqnarray}
The matrix $C$ is the charge conjugation matrix for Dirac spinors
with $C=i\gamma^{2}\gamma^{0}$, and the transpose is on the spinor
matrix indices.  Under the heavy quark spin transformation $S$ and
light quark $SU(3)$ flavor symmetry $U$,
\begin{equation}
\label{11}H^{(\overline{Q})}_a\rightarrow
U_{ab}H^{(\overline{Q})}_bS^{\dagger},~~~S^{(\overline{Q})}_a\rightarrow
U_{ab}S^{(\overline{Q})}_bS^{\dagger},~~~T^{(\overline{Q})}_{a\mu}\rightarrow
U_{ab}T^{(\overline{Q})}_{b\mu}S^{\dagger}
\end{equation}
Similarly the hermitian conjugate fields are defined by
\begin{equation}
\label{12}\overline{H}^{\,(\overline{Q})}_a=\gamma_0H^{(\overline{Q})\dagger}_a\gamma_0,~~~\overline{S}^{\,(\overline{Q})}_a=\gamma_0S^{(\overline{Q})\dagger}_a\gamma_0,~~~\overline{T}^{\,(\overline{Q})}_{a\mu}=\gamma_0T^{(\overline{Q})\dagger}_{a\mu}\gamma_0
\end{equation}
Under the symmetry transformation $S$ and $U$
\begin{equation}
\label{13}\overline{H}^{\,(\overline{Q})}_a\rightarrow
S\,\overline{H}^{\,(\overline{Q})}_b\,U^{\dagger}_{ba},~~~\overline{S}^{\,(\overline{Q})}_a\rightarrow
S\,\overline{S}^{\,(\overline{Q})}_b\,U^{\dagger}_{ba},~~~\overline{T}^{\,(\overline{Q})}_{a\mu}\rightarrow
S\overline{T}^{\,(\overline{Q})}_{b\mu}\,U^{\dagger}_{ba}
\end{equation}
For the system including both heavy meson and heavy anti-meson field
in the HM$\chi$PT, the total effective Lagrangian should be
invariant under the charge conjugation transformation. The
interaction between the pseudo-Goldstone bosons and the meson
containing one heavy anti-quark can be obtained from Eq.(\ref{7}) by
applying the charge conjugation operator
\begin{eqnarray}
\nonumber&&{\cal L}'_{\rm
P}=ig\langle\overline{H}^{\,(\overline{Q})}_a{\cal
A}\!\!\!\slash_{ab}\gamma_5H^{(\overline{Q})}_b\rangle+ik\langle\overline{T}^{\,(\overline{Q})\mu}_a{\cal
A}\!\!\!\slash_{ab}\gamma_5T^{(\overline{Q})}_{b\mu}\rangle+i\tilde{k}\langle\overline{S}^{\,(\overline{Q})}_a{\cal
A}\!\!\!\slash_{ab}\gamma_5S^{(\overline{Q})}_b\rangle+\Big[ih\langle\overline{H}^{\,(\overline{Q})}_a{\cal
A}\!\!\!\slash_{ab}\gamma_5S^{(\overline{Q})}_b\rangle\\
\label{14}&&+i\tilde{h}\langle
\overline{S}^{\,(\overline{Q})}_a{\cal A}_{\mu
ab}\gamma_5T^{(\overline{Q})\mu}_b\rangle+i
\frac{h_1}{\Lambda_{\chi}}\langle\overline{H}^{\,(\overline{Q})}_a({\cal
A}\!\!\!\slash\stackrel{\leftarrow}{D_{\mu}'})_{ab}\gamma_5T^{(\overline{Q})\mu}_b\rangle+i\frac{h_2}{\Lambda_{\chi}}\langle\overline{H}^{\,(\overline{Q})}_a({\cal
A}_{\mu}\stackrel{\leftarrow}{D\!\!\!\slash'})_{ab}\gamma_5T^{(\overline{Q})\mu}_b\rangle+h.c.\Big]
\end{eqnarray}
where $D'_{\mu}=\partial_{\mu}-{\cal V}_{\mu}$. After expanding the
effective Lagrangian in Eq.(\ref{7}) and Eq.(\ref{14}) to the
leading order of pseudo-Goldstone field, we can obtain the following
effective interactions, which is needed in our work
\begin{eqnarray}
\nonumber&&{\cal L}_{\rm DD^{*}P}=g_{\rm DD^{*}P}{\rm
D}_b(\partial_{\mu}{\cal M})_{ba}{\rm D}^{*\mu\dagger}_a+g_{\rm
DD^{*}P}{\rm D}^{*\mu}_{b}(\partial_{\mu}{\cal M})_{ba}{\rm
D}^{\dagger}_a+g_{\rm\overline{D}\,\overline{D}^{*}P}\,\overline{\rm
D}^{*\mu\dagger}_a(\partial_{\mu}{\cal M})_{ab}\overline{\rm
D}_b+g_{\rm\overline{D}\,\overline{D}^{*}P}\,\overline{\rm
D}^{\dagger}_a(\partial_{\mu}{\cal M})_{ab}\overline{\rm
D}^{*\mu}_b\\
\nonumber&&{\cal L}_{\rm D_0D_1P}=g_{\rm D_0D_1P}{\rm
D}^{\mu}_{1b}(\partial_{\mu}{\cal M})_{ba}{\rm
D}^{\dagger}_{0a}+g_{\rm\overline{D}_0\overline{D}_1P}\,\overline{\rm
D}^{\dagger}_{0a}(\partial_{\mu}{\cal M})_{ab}\overline{\rm
D}^{\mu}_{1b}+h.c.\\
\nonumber&&{\cal L}_{\rm DD_0P}=ig_{\rm DD_0P}({\rm
D}_{0b}\stackrel{\leftrightarrow}{\partial}_{\mu}{\rm
D}^{\dagger}_a)\partial_{\mu}{\cal
M}_{ba}+ig_{\rm\overline{D}\,\overline{D}_0P}(\overline{\rm
D}_{0b}\stackrel{\leftrightarrow}{\partial}_{\mu}\overline{\rm
D}^{\dagger}_a)\partial^{\mu}{\cal M}_{ab}+h.c\\
\nonumber&&{\cal L}_{\rm D^{*}D_1P}=g_{\rm D^{*}D_1P}\,\Big[3{\rm
D}^{\mu}_{1b}(\partial_{\mu}\partial_{\nu}{\cal M})_{ba}{\rm
D}^{*\nu\dagger}_a-{\rm
D}^{\mu}_{1b}(\partial^{\nu}\partial_{\nu}{\cal M})_{ba}{\rm
D}^{*\dagger}_{a\mu}+\frac{1}{\rm
M_{D^{*}}M_{D_1}}\partial^{\nu}{\rm
D}^{\mu}_{1b}(\partial_{\nu}\partial_{\tau}{\cal
M})_{ba}\partial^{\tau}{\rm D}^{*\dagger}_{a\mu}\Big]\\
\label{15}&&+g_{\rm\overline{D}^{*}\overline{D}_1P}\,\Big[3\overline{\rm
D}^{*\mu\dagger}_a(\partial_{\mu}\partial_{\nu}{\cal
M})_{ab}\overline{\rm D}^{\nu}_{1b}-\overline{\rm
D}^{*\mu\dagger}_a(\partial^{\nu}\partial_{\nu}{\cal
M})_{ab}\overline{\rm D}_{1b\mu}+\frac{1}{\rm
M_{D^{*}}M_{D_1}}\partial^{\nu}\overline{\rm
D}^{*\mu\dagger}_a(\partial_{\nu}\partial_{\tau}{\cal
M})_{ab}\partial^{\tau}\overline{\rm D}_{1b\mu}\Big]+h.c.
\end{eqnarray}
In the chiral and heavy quark limit, the above coupling constants
are
\begin{eqnarray}
\nonumber&&g_{\rm
DD^{*}P}=-g_{\rm\overline{D}\,\overline{D}^{*}P}=-\frac{2g}{f_{\pi}}\sqrt{\rm
M_{\rm D}M_{\rm D^{*}}}\\
\nonumber&&g_{\rm
D_0D_1P}=g_{\rm\overline{D}_0\overline{D}_1P}=-\frac{2\sqrt{6}}{3}\frac{\tilde{h}}{f_{\pi}}\sqrt{\rm
M_{D_0}M_{D_1}}\\
\nonumber&&g_{\rm
DD_0P}=g_{\rm\overline{D}\,\overline{D}_0P}=-\frac{h}{f_{\pi}}\\
\label{19}&&g_{\rm
D^{*}D_1P}=g_{\rm\overline{D}^{*}\overline{D}_1P}=-\frac{\sqrt{6}}{3}\,\frac{h_1+h_2}{\Lambda_{\chi}f_{\pi}}\sqrt{\rm
M_{D^{*}}M_{D_1}}
\end{eqnarray}
We would like to stress that the ${\rm DD^{*}P}$ coupling constant
is the negative of the ${\rm\overline{D}\,\overline{D}^{*}P}$
coupling constant, because of the phase convention for charge
conjugation chosen in Eq.(\ref{9}). The effective Lagrangian between
$\sigma$ and heavy meson(anti-meson) are \cite{Bardeen:2003kt}
\begin{eqnarray}
\nonumber&&{\cal L}_{\sigma}=g_{\sigma}\langle
H^{(Q)}_a\sigma\overline{H}^{\,({Q})}_a\rangle+g'_{\sigma}\langle
S^{(Q)}_a\sigma\overline{S}^{\,(Q)}_a\rangle+g''_{\sigma}\langle
T^{(Q)\mu}_a\sigma\overline{T}^{\,({Q})}_{a\mu}\rangle+\Big[\frac{h_{\sigma}}{f_{\pi}}\langle
S^{(Q)}_a\gamma^{\mu}(\partial_{\mu}\sigma)\overline{H}^{\,({Q})}_a\rangle\\
\nonumber&&+\frac{h'_{\sigma}}{f_{\pi}}\langle
T^{(Q)\mu}_a(\partial_{\mu}\sigma)\overline{H}^{\,(Q)}_a\rangle+h.c.\Big]+g_{\sigma}\langle\overline{H}^{\,(\overline{Q})}_a\sigma
H^{(\overline{Q})}_a\rangle+g'_{\sigma}\langle\overline{S}^{\,(\overline{Q})}_a\sigma
S^{(\overline{Q})}_a\rangle+g''_{\sigma}\langle\overline{T}^{\,(\overline{Q})\mu}_a\sigma
T^{(\overline{Q})}_{a\mu}\rangle\\
\label{20}&&+\Big[-\frac{h_{\sigma}}{f_{\pi}}\langle\overline{H}^{\,(\overline{Q})}_a\gamma^{\mu}(\partial_{\mu}\sigma)S^{(\overline{Q})}_a\rangle+\frac{h'_{\sigma}}{f_{\pi}}\langle
\overline{H}^{\,(\overline{Q})}_a(\partial_{\mu}\sigma)T^{(\overline{Q})\mu}_b\rangle+h.c.\Big]
\end{eqnarray}
The coupling constants are estimated as follows\cite{Bardeen:2003kt}
\begin{equation}
\label{add1}g_{\sigma}=-\frac{g_{\pi}}{2\sqrt{6}},~~~g'_{\sigma}=-\frac{g_{\pi}}{2\sqrt{6}},~~~h_{\sigma}=\frac{g_A}{\sqrt{3}}
\end{equation}
where $g_{\pi}=3.73$ and $g_{A}=0.6$. As in Ref.\cite{Liu:2008xz},
we take $|g''_{\sigma}|=|g_{\sigma}|$ and
$|h'_{\sigma}|=|h_{\sigma}|$ approximately when performing the
numerical analysis. Expanding the Lagrangian ${\cal L}_{\sigma}$, we
get the interactions associated with $\sigma$
\begin{eqnarray}
\nonumber&&{\cal L}_{\rm DD\sigma}=g_{\rm DD\sigma}\;{\rm D}_a{\rm
D}^{\dagger}_a\sigma+g_{\rm\overline{D}\,\overline{D}\sigma}\,\overline{\rm
D}_a\overline{\rm D}^{\dagger}_a\sigma\\
\nonumber&&{\cal L}_{\rm {D}_1{D}_1{\sigma}}=g_{\rm D_1D_1\sigma}\;
{\rm D}^{\mu}_{1a}{\rm D}^{\dagger}_{1a\mu}\sigma+ g_{\rm
\overline{D}_1\overline{D}_1\sigma}\;\overline{\rm
D}^{\mu}_{1a}\overline{\rm D}^{\dagger}_{1a\mu}\sigma \\
\nonumber&&{\cal L}_{\rm{D}{D}_1{\sigma}}= g_{\rm{D}{D}_1{\sigma}}
{\rm D}^{\mu}_{1a}{\rm D}^{\dagger}_a\partial_{\mu}\sigma+
g_{\rm\overline{D}\,\overline{D}_1{\sigma}}
\overline{\rm D}^{\mu}_{1a}\overline{\rm D}^{\dagger}_a\partial_{\mu}\sigma+h.c.\\
\nonumber&&{\cal L}_{\rm {D}^{*}{D}^{*}{\sigma}}=g_{\rm
D^{*}D^{*}\sigma}\, {\rm D}^{*\mu}_a{\rm D}^{*\dagger}_{a\mu}\sigma+
g_{\rm \overline{D}^{\,*}\overline{D}^{\,*}\sigma}\,
\overline{\rm D}^{*\mu}_a\overline{\rm D}^{*\dagger}_{a\mu}\sigma\\
\nonumber&&{\cal L}_{\rm{D}_0{D}_0{\sigma}}=g_{\rm D_0D_0\sigma}\,
{\rm D}_{0a}{\rm D}^{\dagger}_{0a}\sigma +g_{\rm
\overline{D}_0\overline{D}_0\sigma}\,\overline{\rm
D}_{0a}\overline{\rm D}^{\dagger}_{0a}\sigma\\
\label{21}&&{\cal L}_{\rm{D}^{*}{D}_0{\sigma}}=g_{\rm
D^{*}D_0\sigma}\;{\rm D}_{0a}{\rm
D}^{*\mu\dagger}_a\partial_{\mu}\sigma+ g_{\rm
\overline{D}^{*}\overline{D}_0\sigma}\; \overline{\rm
D}_{0a}\overline{\rm D}^{*\mu\dagger}_a\partial_{\mu}\sigma+h.c.
\end{eqnarray}
The relevant coupling constants are
\begin{eqnarray}
\nonumber&&g_{\rm DD\sigma}=g_{\rm
\overline{D}\,\overline{D}\,\sigma}=-2g_{\sigma}{\rm M_D}\\
\nonumber&&g_{\rm D_1D_1\sigma}=g_{\rm
\overline{D}_1\overline{D}_1\,\sigma}=-2g''_{\sigma}{\rm M_{D_1}}\\
\nonumber&&g_{\rm
{D}{D}_1{\sigma}}=g_{\rm\overline{D}\,\overline{D}_1\,{\sigma}}=-\frac{2\sqrt{6}}{3}\,\frac{h'_{\sigma
}}{f_{\pi}}\sqrt{\rm M_DM_{D_1}}\\
\nonumber&&g_{\rm D^{*}D^{*}\sigma}=g_{\rm\overline{D}^{*}\overline{D}^{*}\sigma}=2g_{\sigma}{\rm M_{D^{*}}}\\
\nonumber&&g_{\rm D_0D_0\sigma}=g_{\rm
\overline{D}_0\overline{D}_0\sigma}=2g'_{\sigma}{\rm M_{D_0}}\\
\label{22}&&g_{\rm D^{*}D_0 \sigma}=-g_{\rm
\overline{D}^{*}\overline{D}_0\,\sigma}=-\frac{2h_{\sigma}}{f_{\pi}}\sqrt{\rm
M_{D^{*}}M_{D_0}}
\end{eqnarray}

\section{Converting the T-matrix into the effective potential}
The T-matrix for $A({\bf p_1})B({\bf p_2})\rightarrow C({\bf
p'_1})D({\bf p'_2})$ scattering process can be represented by an
equivalent Born-order potential operator $V_{bn}({\bf
r_1-r_2},\nabla_1,\nabla_2)$ between pointlike particles, the
definition of this potential operator is \cite{Barnes:1982eg,book}
\begin{eqnarray}
\nonumber&&\delta^{3}({\bf p'_1+p'_2-p_1-p_2}){\rm T}_{fi}({\bf
p_1},{\bf p_2},{\bf p'_1},{\bf p'_2})\\
\label{23}&&=\frac{1}{(2\pi)^3}\int\int d^3{\bf r_1}d^3{\bf
r_2}e^{-i({\bf p'_1\cdot r_1}+{\bf p'_2\cdot r_2})}V_{bn}({\bf
r_1-r_2},\nabla_1,\nabla_2)e^{i({\bf p_1\cdot r_1}+{\bf p_2\cdot
r_2})}
\end{eqnarray}
where ${\rm T}_{fi}({\bf p_1},{\bf p_2},{\bf p'_1},{\bf p'_2})$ is
the T-matrix for the process $A({\bf p_1})B({\bf p_2})\rightarrow
C({\bf p'_1})D({\bf p'_2})$. In general, ${\rm T}_{fi}$ depends on
all the involved momentum ${\bf p_1}$, ${\bf p_2}$, ${\bf p'_1}$ and
${\bf p'_2}$. For convenience, we introduce
\begin{eqnarray}
\label{24}{\bf P_1\equiv\frac{1}{2}(p_1+p'_1)},~~~{\bf
P_2\equiv\frac{1}{2}(p_2+p'_2)},~~~{\bf q\equiv p'_1-p_1\equiv
p_2-p'_2}
\end{eqnarray}
In the center of mass frame ${\bf P_1=-P_2}$. The amplitude ${\rm
T}_{fi}$ can be expanded as a power series in ${\rm P}_{1i}$ and
${\rm P}_{2i}$
\begin{equation}
\label{25}{\rm T}_{fi}({\bf p_1},{\bf p_2},{\bf p'_1},{\bf
p'_2})={\rm T}^{(0)}({\bf q})+{\rm T}^{(1,0)}_i({\bf q}){\rm
P}_{1i}+{\rm T}^{(0,1)}_i({\bf q}){\rm P}_{2i}+{\rm
T}^{(1,1)}_{ij}({\bf q}){\rm P}_{1i}{\rm P}_{2j}+...
\end{equation}
This procedure produces the full Breit-Fermi Hamiltonian when it is
applied to the photon exchanged electron-electron scattering
amplitude expanded to ${\cal O}({\rm P}^2)$. The leading term ${\rm
T}^{(0)}({\bf q})$ is a function of ${\bf q}$ only, its Fourier
transformation gives us a local potential $V({\bf r})$ that is a
function of ${\bf r_1-r_2\equiv r}$ only. The relation between ${\rm
T}^{(0)}({\bf q})$ and $V({\bf r})$ is
\begin{equation}
\label{26}V({\bf r})=\frac{1}{(2\pi)^3}\int d^3{\bf q}\,{\rm
T}^{(0)}({\bf q})e^{i{\bf q\cdot r}}
\end{equation}
For the higher terms of the T-matrix expansion, ${\rm P}_{1i}$ and
${\rm P}_{2i}$ are replaced by left- and right-gradients in the
equivalent potential operator defined implicitly by
Eq.(\ref{23})\cite{Barnes:1982eg}. Following this systematic
procedure, we can convert a general T-matrix into an equivalent
potential operator. In this work, we obtain the local potential by
Fourier transforming the leading terms ${\rm T}^{(0)}({\bf q})$ of
the scattering amplitude ${\rm T}_{fi}$, which is common in
potential model\cite{Barnes:1999hs,Barnes:2000hu}.

Since the propagators are off-shell, we introduce form factor at
each vertex when writing out the scattering amplitude, the usual
form factor is expressed as\cite{Tornqvist:1993ng,Locher:1993cc}
\begin{equation}
\label{27}F(q)=\frac{\Lambda^2-m^2}{\Lambda^2-q^2}
\end{equation}
where $\Lambda$ is an adjustable constant within a reasonable range
of 1-2 GeV, which models the off-shell effects at the vertices due
to the internal structure of the meson. $m$ and $q$ are the mass and
the four momentum of the exchanged meson respectively.

\section{The effective potentials related with Y(4260) and ${\rm Z^{+}_2(4250)}$}

\begin{center}
\begin{figure}[hptb]
\includegraphics*[100pt,335pt][500pt,730pt]{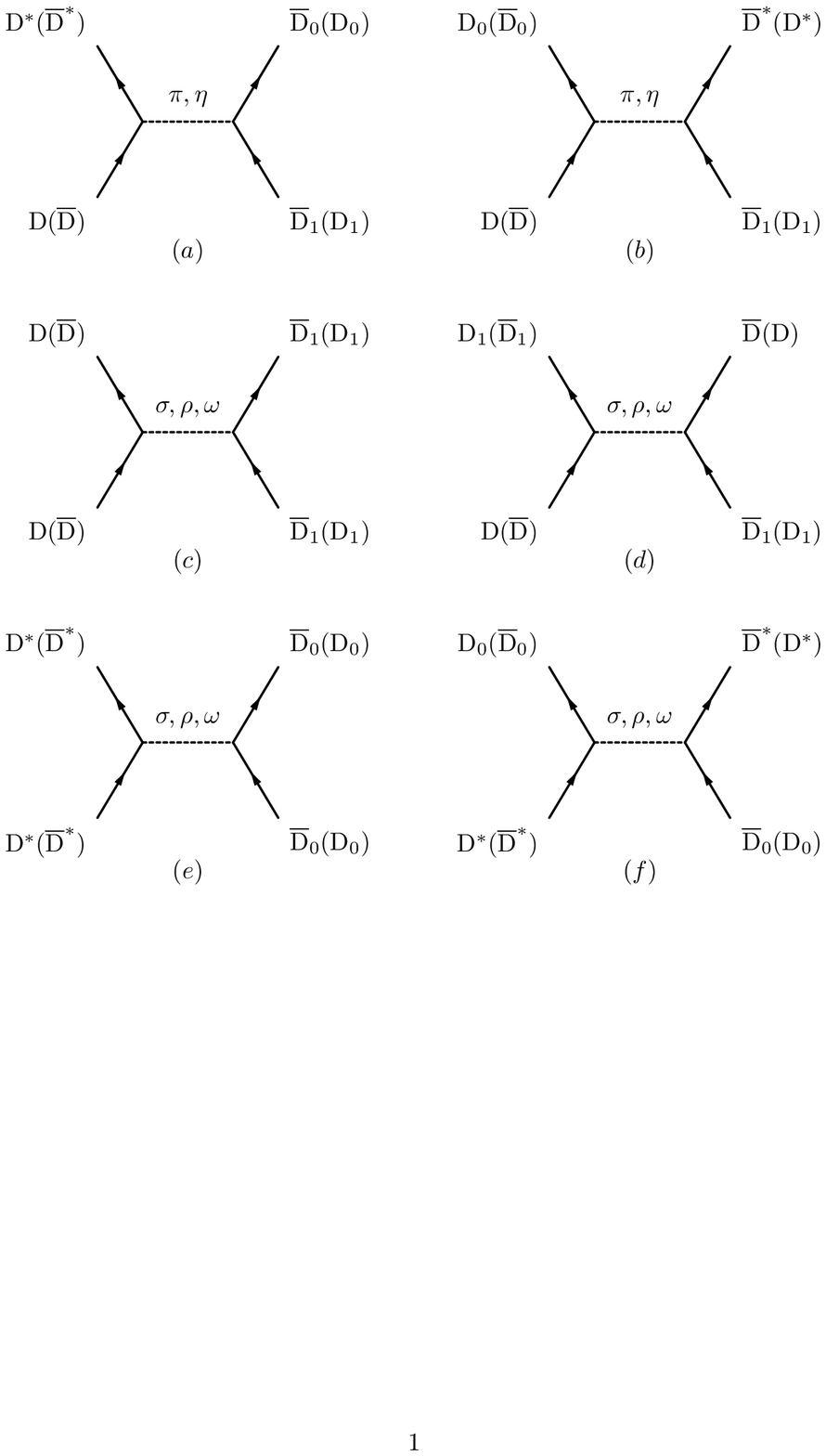}
\caption{\label{scatter-diagram}The scattering diagrams with
pseudoscalars $\pi$, $\eta$ exchange, $\sigma$ exchange, and light
vector mesons $\rho$, $\omega$ exchange.}
\end{figure}
\end{center}

Recently, the meson exchange model based on the HM$\chi$PT has been
used to study possible heavy flavor
molecule\cite{liu-3872,Liu:2008xz}. In this section, we will follow
the general procedure shown above to derive the effective potential
associated with Y(4260) and ${\rm Z^{+}_2(4250)}$ in the framework
of HM$\chi$PT. From the effective interaction in Eq.(\ref{15}) and
Eq.(\ref{21}), we can write down the corresponding scattering
amplitude for each diagram, including the form factor at each
vertex. Then we get the equivalent potential in momentum space
following the general formalism presented in section III. Finally we
make Fourier transformation to derive the potentials in coordinate
space. Because of parity conservation, pseudoscalar $\pi$ and $\eta$
exchange only contributes to the off-diagonal interaction, whereas
$\sigma$ exchange and light vector mesons $\rho$, $\omega$ exchange
result in diagonal interaction only. The corresponding scattering
diagrams are shown in Fig.\ref{scatter-diagram}.
Under the ansatz of Y(4260) as a ${\rm D_1D}$ or ${\rm D_0D^*}$
hadronic molecule, we can write down its flavor wavefunction
\begin{eqnarray}
\nonumber&&{\rm
|Y(4260)\rangle=\frac{1}{2}[|D^0_1\overline{D}^{\,0}\rangle+|D^{+}_1D^-\rangle-|D^0\overline{D}^{\,0}_1\rangle-|D^+D^-_1\rangle]}\\
&&{\rm
\label{28}|Y'(4260)\rangle=\frac{1}{2}[|D^{0}_0\overline{D}^{*0}\rangle+|D^{+}_0D^{*-}\rangle+|D^{*0}\overline{D}^{\,0}_0\rangle+|D^{*+}D^-_0\rangle]}
\end{eqnarray}
We stress that the phase convention under charge conjugation is
consistent with Eq.(\ref{9}). In the same way, the flavor
wavefunction of ${\rm Z^{+}_2(4250)}$ is
\begin{eqnarray}
\nonumber&& {\rm
|{\rm Z^{+}_2(4250)}\rangle=\frac{1}{\sqrt{2}}[|D^{+}_1\overline{D}^{\,0}\rangle+|D^{+}\overline{D}^{\,0}_1\rangle]}\\
\label{29}&&{\rm |{\rm
Z'^{+}_2(4250)}\rangle=\frac{1}{\sqrt{2}}[|D^{+}_0\overline{D}^{*0}\rangle-|D^{*+}\overline{D}^{\,0}_0\rangle]}
\end{eqnarray}
In this case, its quantum number are ${\rm
I^{G}(J^{P})=1^{-}(1^{-})}$. Following the procedure discussed
above, we can calculate the effective potential in momentum space,
it is a lengthy and tedious calculation.

\begin{figure}[hptb]
\begin{center}
\begin{tabular}{cc}
\includegraphics[height=4cm]{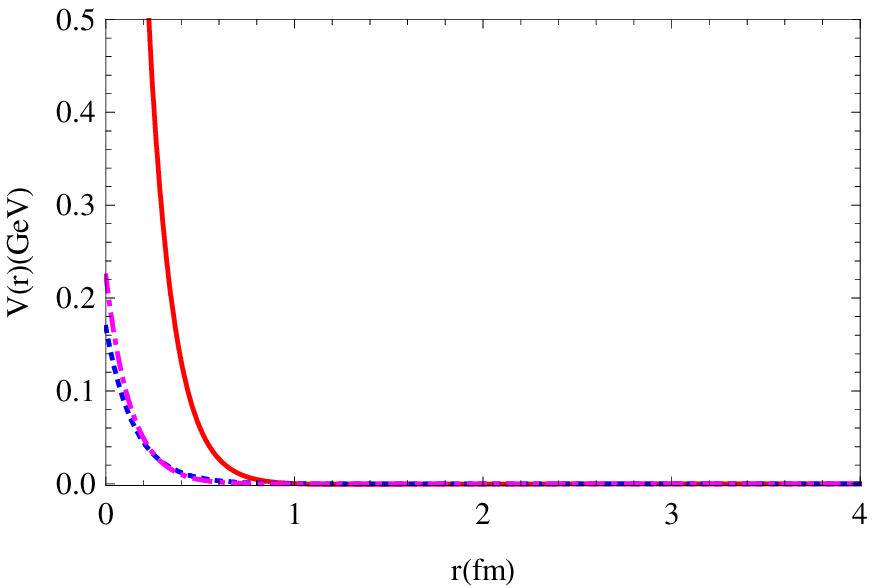}&\includegraphics[height=4cm]{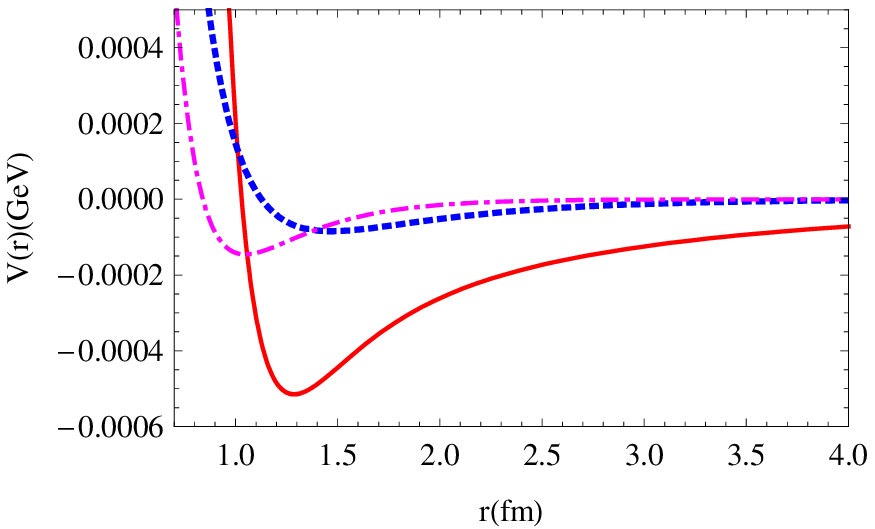}\\
(a)&(b)\\\\
\includegraphics[height=4cm]{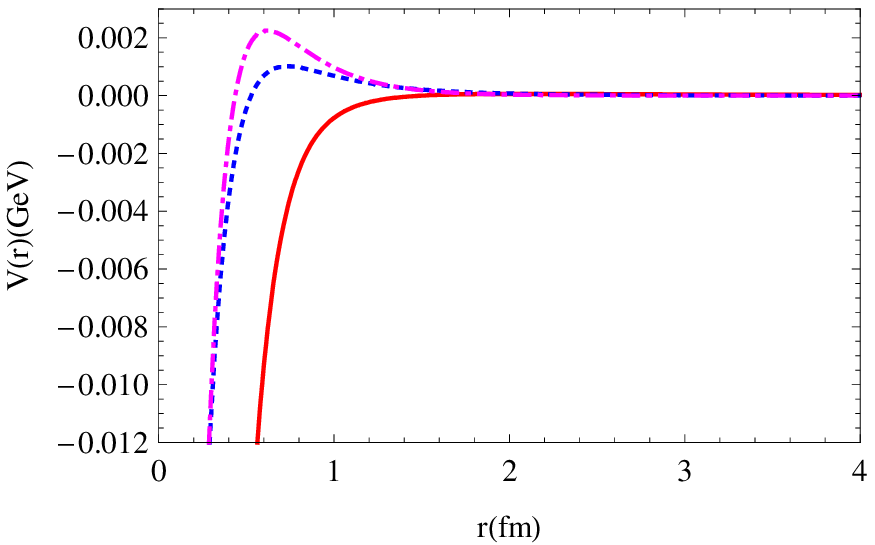}&\includegraphics[height=4cm]{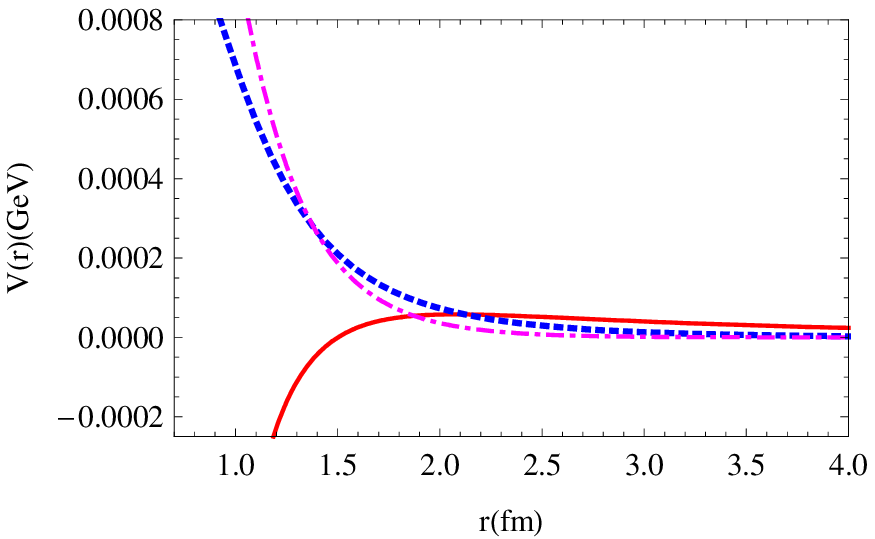}\\
(c)&(d)
\end{tabular}
\caption{\label{all-component-potential}(color online)The effective
potential for the ${\rm D_1D}$ and ${\rm D_0D^{*}}$ system from
pseudoscalar $\pi$, $\eta$ exchange and scalar $\sigma$ exchange.
Solid line represents the non-diagonal potential $V_{12}(r)$ (or
$V_{21}(r)$), short dashed and dash dotted lines respectively
correspond to the diagonal potential $V_{11}(r)$ and $V_{22}(r)$.
(a) and (b) are related with Y(4260), and (b) shows the long range
behavior of the potential. (c) and (d) are related with ${\rm
Z^{+}_2(4250)}$, and (d) is the long rang shape of the potential.}
\end{center}
\end{figure}

For Y(4260), the exchange potential in momentum space is
\begin{eqnarray}
\nonumber&&V_{12}({\bf q})=V_{21}({\bf
q})=\frac{\sqrt{6}}{6}\frac{g\tilde{h}}{f^2_{\pi}}\Big(\frac{\Lambda^2-m^2_{\pi}}{{\bf
q}^2+X^2_1}\Big)^2\frac{{\bf q}^2}{{\bf
q}^2+\mu^2_1}+\frac{\sqrt{6}}{54}\frac{g\tilde{h}}{f^2_{\pi}}\Big(\frac{\Lambda^2-m^2_{\eta}}{{\bf
q}^2+X^2_1}\Big)^2 \frac{{\bf
q}^2}{{\bf q}^2+\mu^2_2}\\
\nonumber&&V_{11}({\bf q})=\Big(\frac{\Lambda^2-m^2_{\sigma}}{{\bf
q}^2+\Lambda^2}\Big)^2 \frac{g_{\sigma}g''_{\sigma}}{{\bf
q}^2+m^2_{\sigma}}+\frac{2h'^{\,2}_{\sigma}}{9f^2_{\pi}}\Big(\frac{\Lambda^2-m^2_{\sigma}}{{\bf
q}^2+X^2_2}\Big)^2 \frac{{\bf
q}^2}{{\bf q}^2+\mu^2_3}\\
\label{30}&&V_{22}({\bf q})=\Big(\frac{\Lambda^2-m^2_{\sigma}}{{\bf
q}^2+\Lambda^2}\Big)^2\frac{g_{\sigma}g'_{\sigma}}{{\bf
q}^2+m^2_{\sigma}}+\frac{h^2_{\sigma}}{3f^2_{\pi}}\Big(\frac{\Lambda^2-m^2_{\sigma}}{{\bf
q}^2+X^2_3}\Big)^2\frac{{\bf q}^2}{{\bf q}^2+\mu^2_4}
\end{eqnarray}
where we have include the monopole form factor in Eq.(\ref{27}) to
regularize the potential. The diagonal potential $V_{11}({\bf q})$
and  $V_{22}({\bf q})$ is induced by $\sigma$ exchange, and the
non-diagonal potential $V_{12}({\bf q})$ ( or $V_{21}({\bf q})$ )
arises from the pseudo-Goldstone bosons $\pi$ and $\eta$ exchange.
The effective potential from $\rho$, $\omega$ exchange is shown in
the Appendix.  The potential for ${\rm Z^{+}_2(4250)}$ in momentum
space is
\begin{eqnarray}
\nonumber&&V_{12}({\bf q})=V_{21}({\bf
q})=-\frac{\sqrt{6}}{18}\frac{g\tilde{h}}{f^2_{\pi}}\Big(\frac{\Lambda^2-m^2_{\pi}}{{\bf
q}^2+X^2_1}\Big)^2\frac{{\bf q}^2}{{\bf
q}^2+\mu^2_1}+\frac{\sqrt{6}}{54}\frac{g\tilde{h}}{f^2_{\pi}}\Big(\frac{\Lambda^2-m^2_{\eta}}{{\bf
q}^2+X^2_1}\Big)^2 \frac{{\bf
q}^2}{{\bf q}^2+\mu^2_2}\\
\nonumber&&V_{11}({\bf q})=\Big(\frac{\Lambda^2-m^2_{\sigma}}{{\bf
q}^2+\Lambda^2}\Big)^2 \frac{g_{\sigma}g''_{\sigma}}{{\bf
q}^2+m^2_{\sigma}}-\frac{2h'^{\,2}_{\sigma}}{9f^2_{\pi}}\Big(\frac{\Lambda^2-m^2_{\sigma}}{{\bf
q}^2+X^2_2}\Big)^2 \frac{{\bf
q}^2}{{\bf q}^2+\mu^2_3}\\
\label{31}&&V_{22}({\bf q})=\Big(\frac{\Lambda^2-m^2_{\sigma}}{{\bf
q}^2+\Lambda^2}\Big)^2\frac{g_{\sigma}g'_{\sigma}}{{\bf
q}^2+m^2_{\sigma}}-\frac{h^2_{\sigma}}{3f^2_{\pi}}\Big(\frac{\Lambda^2-m^2_{\sigma}}{{\bf
q}^2+X^2_3}\Big)^2\frac{{\bf q}^2}{{\bf q}^2+\mu^2_4}
\end{eqnarray}
The various parameters appearing in the above formulas are defined
as follows.
\begin{eqnarray}
\nonumber&&X^2_1={\Lambda^2-({\rm M}_{\rm D^{*}}-{\rm M}_{\rm
D})({\rm M}_{\rm D_1}-{\rm M}_{\rm
D_0})}\\
\nonumber&&X^2_2=\Lambda^2-({\rm M}_{\rm D_1}-{\rm M}_{\rm D})^2\\
\nonumber&&X^2_3=\Lambda^2-({\rm M}_{\rm D_0}-{\rm M}_{\rm
D^{*}})^2\\
\nonumber&&\mu^2_1=m^2_{\pi}-({\rm M}_{\rm D^{*}}-{\rm M}_{\rm
D})({\rm M}_{\rm D_1}-{\rm M}_{\rm
D_0})\\
\nonumber&&\mu^2_2=m^2_{\eta}-({\rm M}_{\rm D^{*}}-{\rm M}_{\rm
D})({\rm M}_{\rm D_1}-{\rm M}_{\rm
D_0})\\
\nonumber&&\mu^2_3=m^2_{\sigma}-({\rm M}_{\rm D_1}-{\rm M}_{\rm
D})^2\\
\label{32}&&\mu^2_4=m^2_{\sigma}-({\rm M}_{\rm D_0}-{\rm M}_{\rm
D^{*}})^2
\end{eqnarray}
After performing Fourier transformation, we obtain the potential
forms in configuration space, For Y(4260), the potential in
coordinate space is
\begin{eqnarray}
\nonumber&&V_{12}(r)=V_{21}(r)=\frac{\sqrt{6}}{6}\frac{g\tilde{h}}{f^2_{\pi}}Z(\Lambda,X_1,\mu_1,m_{\pi},r)+\frac{\sqrt{6}}{54}\frac{g\tilde{h}}{f^2_{\pi}}Z(\Lambda,X_1,\mu_2,m_{\eta},r)\\
\nonumber&&V_{11}(r)=g_{\sigma}g''_{\sigma}H(\Lambda,m_{\sigma},r)+\frac{2h'^{2}_{\sigma}}{9f^2_{\pi}}Z(\Lambda,X_2,\mu_3,m_{\sigma},r)\\
\label{33}&&V_{22}(r)=g_{\sigma}g'_{\sigma}H(\Lambda,m_{\sigma},r)+\frac{h^2_{\sigma}}{3f^2_{\pi}}Z(\Lambda,X_3,\mu_4,m_{\sigma},r)
\end{eqnarray}
The potential in coordinate space for ${\rm Z^{+}_2(4250)}$ is
\begin{eqnarray}
\nonumber&&V_{12}(r)=V_{21}(r)=-\frac{\sqrt{6}}{18}\frac{g\tilde{h}}{f^2_{\pi}}Z(\Lambda,X_1,\mu_1,m_{\pi},r)+\frac{\sqrt{6}}{54}\frac{g\tilde{h}}{f^2_{\pi}}Z(\Lambda,X_1,\mu_2,m_{\eta},r)\\
\nonumber&&V_{11}(r)=g_{\sigma}g''_{\sigma}H(\Lambda,m_{\sigma},r)-\frac{2h'^{2}_{\sigma}}{9f^2_{\pi}}Z(\Lambda,X_2,\mu_3,m_{\sigma},r)\\
\label{34}&&V_{22}(r)=g_{\sigma}g'_{\sigma}H(\Lambda,m_{\sigma},r)-\frac{h^2_{\sigma}}{3f^2_{\pi}}Z(\Lambda,X_3,\mu_4,m_{\sigma},r)
\end{eqnarray}
Here the functions $H(\Lambda,m,r)$ and $Z(\Lambda,X,\mu,m,r)$ are
defined as
\begin{eqnarray}
\nonumber&&H(\Lambda,m,r)=\frac{1}{4\pi}\frac{1}{r}(e^{-mr}-e^{-\Lambda
r})-\frac{\Lambda^2-m^2}{8\pi\Lambda}e^{-\Lambda r}\\
\label{35}&&Z(\Lambda,X,\mu,m,r)=\frac{1}{4\pi}\frac{1}{r}(X^2e^{-Xr}-\mu^2e^{-\mu
r})+\frac{\Lambda^2-m^2}{8\pi}(X-\frac{2}{r})e^{-Xr}
\end{eqnarray}
We take the typical values of the coupling constants
$g\tilde{h}=0.85$, $g_{\sigma}g'_{\sigma}=0.58$,
$g_{\sigma}g''_{\sigma}=0.58$ and $|h_{\sigma}|=|h'_{\sigma}|=0.35$,
and $\Lambda=1.5$GeV is  chosen for an illustration, the variation
of the effective potential with respect to r is shown in
Fig.\ref{all-component-potential}. It is obvious that the magnitude
of the diagonal potentials from $\sigma$ exchange is smaller than
that of the off-diagonal potential from $\pi$ and $\eta$ exchange,
this is mainly because $m_{\pi}$ is small than $m_{\sigma}$.
Moreover, the magnitude of the off-diagonal potential related with
Y(4260) are larger that associated with ${\rm Z^{+}_2(4250)}$, the
latter is about one third of the former. This is consistent with
results from chiral quark model\cite{Tornqvist:1993ng}, consequently
the ${\rm I^{G}(J^{PC})=0^{-}(1^{--})}$ configuration is easier to
bind than the ${\rm I^{G}(J^{P})=1^{-}(1^{-})}$ configuration.

\section{The structures of Y(4260) and ${\rm Z^{+}_2(4250)}$ and the bottom analog}
\subsection{The bound states of the ${\rm D_1D}$ and ${\rm D_0D^*}$ system with the structure of Y(4260) and ${\rm Z^{+}_2(4250)}$ }
With the above effective potential, we shall explore whether there
are bound states with ${\rm I^{G}(J^{PC})=0^-(1^{--})}$ or ${\rm
I^{G}(J^{P})=1^{-}(1^{-})}$ in the ${\rm D_1D}$ and ${\rm D_0D^*}$
system, by means of solving the two channels coupled
Schr$\ddot{\rm{o}}$dinger equation. There are various methods to
integrate the coupled-channel Schr$\ddot{\rm{o}}$dinger equation
numerically. In this work we shall employ two packages
MATSCS\cite{matlab} and FESSDE2.2\cite{fessde} to perform the
numerical calculation so that the results obtained by one program
can be checked by another. The first package is a Matlab software,
and the second is written in Fortran77. Both packages can fastly and
accurately solve the eigenvalue problem for systems of coupled
Schr$\ddot{\rm{o}}$dinger equations, and the results obtained by two
codes are the same within error.

The masses of the involved mesons are taken from PDG\cite{pdg}:
${\rm M_{D}}=1869.3$MeV, ${\rm M_{D^{*}}}=2006.7$MeV, ${\rm
M_{D_1}}=2422$MeV, ${\rm M_{D_0}}=2308$MeV, $m_{\pi}=135$MeV,
$m_{\eta}=547.5$MeV, $m_{\sigma}=600$MeV, $m_{\rho}=775.5$MeV and
$m_{\omega}=782.65$MeV. The effective coupling constants in
HM$\chi$PT have been studied from various phenomenological and
theoretical approaches, and the estimates for $g$, $\tilde{h}$ are
listed in Table \ref{coupling}. It is obvious that there are still
large uncertainties in their values. In the following, we shall
first consider whether one pseudoscalar $\pi$ and $\eta$ exchange
can results in a bound state in the ${\rm D_1D}$ and ${\rm
D_0D^{*}}$ system, then the contribution of $\sigma$ exchange is
included.

\begin{table}[hptb]
\caption{\label{coupling}Summary of theoretical estimates for the
effective coupling $g$ and $\tilde{h}$. }
\bigskip
\begin{tabular}{lllll}\hline \hline
Reference & $~~{ g}$ & {\rm Remark}\\\hline
\cite{Isola:2003fh}&$0.59\pm0.07\pm0.01$&combining the CLEO's results on ${\rm D^{*}}$ decay width \\
\cite{Deandrea:1999pa}&$0.46\pm0.04$&through a constituent
quark-meson model\\
\cite{Fajfer:2006hi}& 0.53& including one loop corrections without
positive parity states\\
\cite{Fajfer:2006hi}& 0.65& including one loop corrections with
positive parity states\\
\cite{Casalbuoni:1996pg}&$0.44\pm0.16$& from QCD sum rule\\
\cite{Colangelo:1994es}&$0.39\pm0.16$& from QCD sum rule\\
\cite{Belyaev:1994zk}&$0.32\pm0.02$&\\
\cite{Falk:1992cx}&0.75&from non-relativistic quark
model\\\hline\hline\\

Reference & $~~{\tilde{h}}$ & {\rm Remark}\\\hline
\cite{Falk:1992cx}&$|\tilde{h}|=0.87$& from non-relativistic quark\\
\cite{Polosa:1999da}&$0.91^{+0.5}_{-0.3}$& in a constituent
quark-meson model in soft pion limit\\\hline\hline
\end{tabular}
\end{table}

The numerical results with only one pseudoscalar exchange are
presented in Table \ref{charm-table1}. For several typical values of
$g\tilde{h}$, we vary the cutoff $\Lambda$ from a small value until
we find a solution which lies below the ${\rm D_1D}$ threshold. Here
the mass ${\rm M}$ is measured with respect to the ${\rm D_1D}$
threshold ${\rm M_{D_1}+M_{D}}\simeq4291.3$ MeV, ${\rm r_{rms}}$ is
the root of mean square radius, and ${\rm R}$ denotes the ratio
between the ${\rm D_1D}$ and ${\rm D_0D^{*}}$ components in the
bound state solutions. By comparing the results with different
$\Lambda$ for the same value of the parameter $g\tilde{h}$, one
notes that the magnitude of M increases with $\Lambda$, whereas the
reverse is true for ${\rm r_{rms}}$ and R. The bound state mass is
sensitive to the parameter $g\tilde{h}$ as well, larger $g\tilde{h}$
is helpful to form a molecular state. From the numerical results in
Table \ref{charm-table1}, we see that one can get a molecular state
consistent with Y(4260), given appropriate value for $g\tilde{h}$
and a reasonable cutoff $\Lambda$ in the range 1-2 GeV. However, the
existence of a bound state ${\rm with~ I^{G}(J^{P})=1^{-}(1^{-})}$
require that the value of $\Lambda$ should be at least larger than 4
GeV. The cutoff parameter $\Lambda$ is a typical hadronic scale,
which is generally expected to be in the range 1-2 GeV. If $\Lambda$
is required to be much larger than 2 GeV in order to form a bound
state, we tend to conclude that such a bound state should not exist.
Therefore, it is not appropriate to assign ${\rm Z^{+}_2(4250)}$ as
a ${\rm D_1D}$ or ${\rm D_0D^{*}}$ molecule, if only the
non-diagonal interaction from $\pi$ and $\eta$ exchange is
considered.

\begin{figure}[hptb]
\begin{center}
\begin{tabular}{cc}
\includegraphics[width=0.4\textwidth]{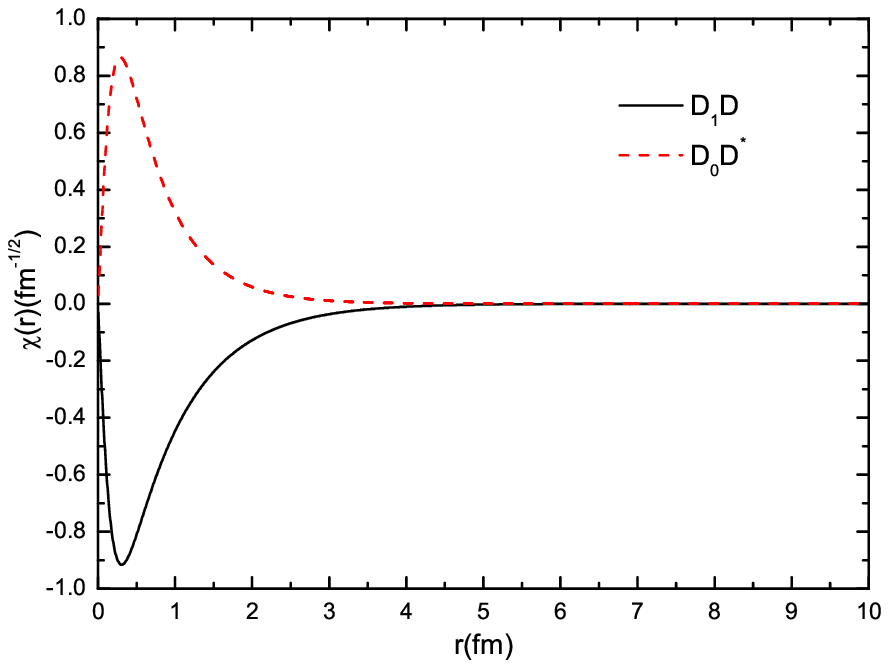}&\includegraphics[width=0.4\textwidth]{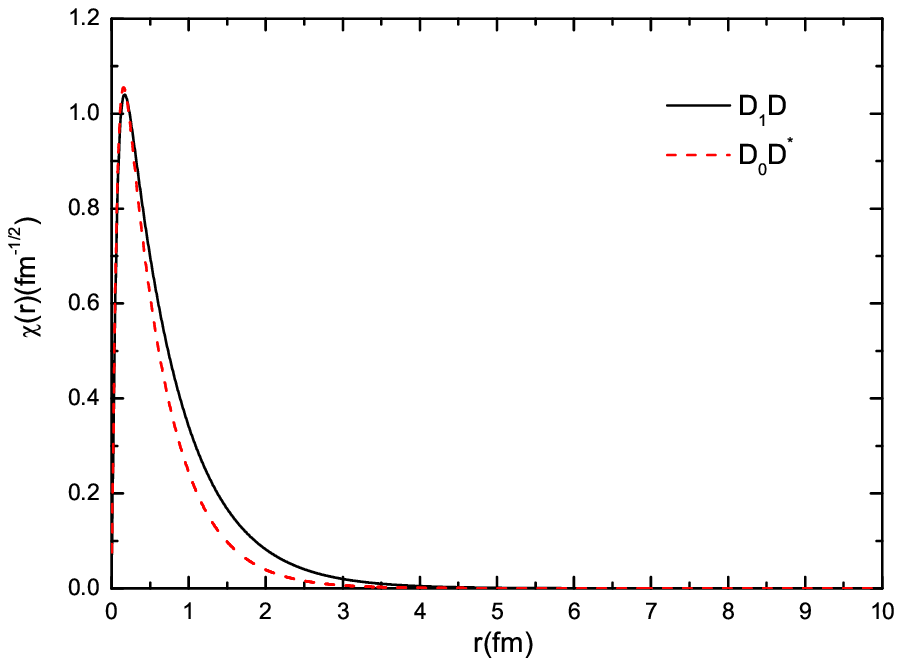}\\
(a)&(b)
\end{tabular}
\caption{\label{wavefunction}(color online) The radial wavefunction
$\chi(r)=rR(r)$ for the molecular states ${\rm with~
I^{G}(J^{PC})=0^{-}(1^{--})}$ and ${\rm I^{G}(J^{P})=1^{-}(1^{-})}$
respectively. (a) corresponds to the former state, and (b) is for
the latter. We have taken $g\tilde{h}=0.85$,
$g_{\sigma}g'_{\sigma}=0.58$, $g_{\sigma}g''_{\sigma}=0.58$ and
$|h_{\sigma}|=h'_{\sigma}=0.35$; $\Lambda$ is chosen to be 1.4 GeV
and 3.4 GeV respectively.}
\end{center}
\end{figure}

Then we include the contribution coming from $\sigma$ exchange,
which leads to only the diagonal interaction. The corresponding
numerical results are shown in Table \ref{charm-table2} and Table
\ref{charm-table3}. The radial wavefunctions $\chi(r)=rR(r)$ for
certain certain parameter values are shown in
Fig.\ref{wavefunction}. The wavefunction corresponding to other
solutions in Table \ref{charm-table2} and \ref{charm-table3} has
similar shape with that in Fig.\ref{wavefunction}. We find that the
$\sigma$ exchange interaction has significant effects, the
variations of M, ${\rm r_{rms}}$ and R with respect to $\Lambda$
have the same pattern as those in the only pseudoscalar exchange
case. Varying the parameters $g\tilde{h}$, $g_{\sigma}g'_{\sigma}$,
$g_{\sigma}g''_{\sigma}$ and $h_{\sigma}$ in the reasonable range
results in large change of the predictions, which indicates that the
results are sensitive to the effective coupling constants. We can
see that large $g\tilde{h}$, negative $g_{\sigma}g'_{\sigma}$ and
$g_{\sigma}g''_{\sigma}$ are favorable to binding the molecular
states. Comparing the results in Table \ref{charm-table1},
\ref{charm-table2} and \ref{charm-table3}, we find that $\sigma$
exchange is against the formation of bound state ${\rm with~
I^{G}(J^{PC})=0^{-}(1^{--})}$, nevertheless, it is beneficial to the
formation of ${\rm I^{G}(J^{P})=1^{-}(1^{-})}$ molecular state. As
for Y(4260), the conclusion reached with only pseudoscalar exchange
remains. Y(4260) could be accommodated as a molecule state for
appropriate effective coupling constants and cutoff. A ${\rm
I^{G}(J^{P})=1^{-}(1^{-})}$ bound state around 4250 MeV requires
$\Lambda$ should be at least 3 GeV, therefore we conclude that the
interpretation of ${\rm Z^{+}_2(4250)}$ as a ${\rm D_1D}$ or ${\rm
D_0D^{*}}$ molecule is disfavored. This conclusion is consistent
with the general observations from chiral quark model. It is found
that the isoscalar channel is easier to bind that the isovector
channel for the same components\cite{Swanson:2006st}.
\subsection{The bottom analog of Y(4260) and ${\rm Z^{+}_2(4250)}$ }
The bottom analog ${\rm Y_{bb}}$ and ${\rm Z^{+}_{bb}}$ respectively
denote the states obtained by replacing both the charm quark and
antiquark with bottom quark and antiquark in Y(4260) and ${\rm
Z^{+}_2(4250)}$. The above calculation can be easily extended to
study these states. The shape of both the diagonal and non-diagonal
potential is similar to that of the charm system, except that the
former is larger than the latter in magnitude. Furthermore, Since
the kinetic energy is greatly reduced because of the heavier mass of
B meson, a molecular state is more easily formed. We choose the same
set of parameters as in the previous section. The numerical results
with only pseudoscalar $\pi$, $\eta$ exchange are shown in Table
\ref{bottom-table1}, and the results with both pseudoscalar and
$\sigma$ exchange are listed in Table \ref{bottom-table2} and
\ref{bottom-table3}. As is expected, the magnitude M of the bottom
analog is larger than that of the corresponding charmed state for
the same parameters. The variation of M, ${\rm r_{rms}}$ and R with
$\Lambda$ is the same as the charm system, large $g\tilde{h}$,
negative $g_{\sigma}g'_{\sigma}$ and $g_{\sigma}g''_{\sigma}$ are
beneficial to molecule formation as well. From the results in Table
\ref{bottom-table1}, \ref{bottom-table2} and \ref{bottom-table3}, we
note that both the bottom analog ${\rm Y_{bb}}$ and ${\rm
Z^{+}_{bb}}$ may exist.

Since Y(4260) has a large branch ratio into $\pi^{+}\pi^{-}J/\psi$,
the bottom analog ${\rm Y_{bb}}$ should be searched for in the
$\pi{+}\pi^{-}\Upsilon$ channel. Specifically, the state ${\rm
Y_{bb}}$ can be searched for at B factories and future Super B
factory via initial state radiation
$e^{+}e^{-}\rightarrow\gamma_{ISR}\pi^{+}\pi^{-}\Upsilon$ or by
$e^{+}e^{-}\rightarrow\pi^{+}\pi^{-}\Upsilon$ direct
scan\cite{Hou:2006kn}. Furthermore, ${\rm Y_{bb}}$ may be searched
for at Tevatron via $p\bar{p}\rightarrow {\rm
Y_{bb}}\rightarrow\pi^{+}\pi^{-}\Upsilon$, and LHC is more
promising. Similarly, for the bottom analog ${\rm Z^{+}_{bb}}$, the
most hopeful discovery channel would be ${\rm Z^{+}_{bb}}\rightarrow
\pi^{+}\chi_{b1}$, where $\chi_{b1}$ is in turn detected by its
decay into $\gamma\Upsilon$\cite{pdg}. Because of the large mass of
this state, it is difficult to produce such state via decay of
certain particle ( i.e., ${\rm Z^{+}_2(4250)}$ is produced in B
decay\cite{Mizuk:2008me} ), consequently large hadron collides such
as Tevatron and LHC are good place to search for this state.

\section{Conclusion and discussions}

In this work, we have performed a dynamical study of Y(4260) and
${\rm Z^{+}_2(4250)}$ simultaneously to see whether they could be
${\rm D_1D}$ or ${\rm D_0D^{*}}$ hadronic molecule. We have employed
the HM$\chi$PT, which combines the heavy quark symmetry and the
chiral symmetry. Since both the heavy meson and heavy anti-meson are
involved, the interaction related with heavy anti-meson has been
included explicitly, and the total effective Lagrangian is invariant
under the charge conjugation transformation.

The off-diagonal interaction from pseudoscalar $\pi$, $\eta$
exchange plays a dominant role, which is a straightforward support
to the off-diagonal interaction mechanism proposed by Swanson and
Close. $\sigma$ exchange leads to only diagonal interaction, its
contribution has been taken into account in this work. We find that
$\sigma$ exchange is not favorable to the formation of molecular
state ${\rm with~ I^{G}(J^{PC})=0^{-}(1^{--})}$, whereas it is
helpful to the binding of molecule ${\rm with~
I^{G}(J^{P})=1^{-}(1^{-})}$. For appropriate value of the effective
coupling constants and a reasonable cutoff $\Lambda$, Y(4260) could
be accommodated as a ${\rm D_1D}$ and ${\rm D_0D^{*}}$ molecule.
However, the existence of a molecule around 4250 MeV ${\rm with~
I^{G}(J^{P})=1^{-}(1^{-})}$ requires that $\Lambda$ should be at
least 3 GeV, given the currently allowed values of the coupling
constants. Consequently, the interpretation of ${\rm Z^{+}_2(4250)}$
as a ${\rm D_1D}$ or ${\rm D_0D^{*}}$ molecule is disfavored. Its
structure should be studied further. Through calculating the masses
of excited heavy tetraquarks with hidden charm in the
diquark-antidiquark picture. the authors in Ref.\cite{Ebert:2008kb}
suggested that ${\rm Z^{+}_2(4250)}$ could be the charged partner of
the $1^{-}$ $1\rm{P}$ state $\rm{S\bar{S}}$ or as the $0^-$
$1\rm{P}$ state of the $\rm{(S\bar{A}\pm \bar{S}A)/\sqrt{2} }$
tetraquark. QCD sum rule analysis for ${\rm Z^{+}_2(4250)}$ is
performed in the Ref.\cite{Wang:2008af}.

The effective potential from vector meson $\rho$, $\omega$ exchange
has been presented analytically. Because of the accidental
coincidence of $m_{\rho}$ and $m_{\omega}$, the contribution from
$\rho$ and $\omega$ exchange almost cancels in the potential related
with ${\rm Z^{+}_2(4250)}$. For Y(4260), the situation is not the
same. A number of effective coupling constants are involved. Because
some of them have not been determined so far, we can not give a
quantitative estimate about the contribution from vector meson
exchange. Qualitatively, it should be smaller than the contribution
coming from pseudoscalar and $\sigma$ exchange in magnitude. It is
necessary and interesting to examine the effect of vector meson
exchange on Y(4260) in future.

The bottom analog of Y(4260) and ${\rm Z^{+}_2(4250)}$ denoted by
${\rm Y_{bb}}$ and ${\rm Z^{+}_{bb}}$ respectively may exist. ${\rm
Y_{bb}}$ can be searched for in
$e^{+}e^{-}\rightarrow\gamma_{ISR}\pi^{+}\pi^{-}\Upsilon$ or by
$e^{+}e^{-}\rightarrow\pi^{+}\pi^{-}\Upsilon$ direct scan. The
direct production of ${\rm Y_{bb}}$ at Tevatron or LHC via
$p\bar{p}\rightarrow {\rm Y_{bb}}\rightarrow\pi^{+}\pi^{-}\Upsilon$
is a hopeful approach as well. For ${\rm Z^{+}_{bb}}$, the most
promising discovery channel is ${\rm
Z^{+}_{bb}}\rightarrow\pi^{+}\chi_{b1}$.

\section*{ACKNOWLEDGEMENTS}
\indent  We acknowledge Prof. Dao-Neng Gao and Prof. Mu-Lin Yan for
very helpful and stimulating discussions, and we are grateful to Dr.
Yan-Rui Liu for useful communications. This work is supported by
China Postdoctoral Science foundation (20070420735).

\begin{appendix}
\section{The potential from light vector mesons $\rho$ and $\omega$ exchange}
The light vector mesons nonet can be introduced by using the hidden
gauge symmetry approach, and the Lagrangian containing these
particles is as
follows\cite{Casalbuoni:1996pg,Casalbuoni:1992gi,Casalbuoni:1992dx}
\begin{eqnarray}
\nonumber&&{\cal L}_V=i\beta\langle H^{(Q)}_bv^{\mu}({\cal
V}_{\mu}-\rho_{\mu})_{ba}\overline{H}^{\,(Q)}_a\rangle+i\lambda\langle
H^{(Q)}_{b}\sigma^{\mu\nu}F_{\mu\nu}(\rho)_{ba}\overline{H}^{\,(Q)}_a\rangle+i\beta_1\langle
S^{(Q)}_bv^{\mu}({\cal V}_{\mu}-\rho_{\mu})_{ba}\overline{S}^{\,(Q)}_a\rangle\\
\nonumber&&+i\lambda_1\langle
S^{(Q)}_b\sigma^{\mu\nu}F_{\mu\nu}(\rho)_{ba}\overline{S}^{\,(Q)}_a\rangle+i\beta_2\langle
T^{(Q)\lambda}_bv^{\mu}({\cal
V}_{\mu}-\rho_{\mu})_{ba}\overline{T}^{\,(Q)}_{a\lambda}\rangle+i\lambda_2\langle
T^{(Q)\lambda}_b\sigma^{\mu\nu}F_{\mu\nu}(\rho)_{ba}\overline{T}^{\,(Q)}_{a\lambda}\rangle\\
\nonumber&&+\Big[i\zeta\langle H^{(Q)}_b\gamma^{\mu}({\cal
V}_{\mu}-\rho_{\mu})_{ba}\overline{S}^{\,(Q)}_a\rangle+i\mu\langle
H^{(Q)}_{b}\sigma^{\lambda\nu}F_{\lambda\nu}(\rho)_{ba}\overline{S}^{\,(Q)}_a\rangle+i\zeta_1\langle
T^{(Q)\mu}_b({\cal V}_{\mu}-\rho_{\mu})_{ba}\overline{H}^{\,(Q)}_a\rangle\\
\label{a1}&&+\mu_{1}\langle
T^{(Q)\mu}_b\gamma^{\nu}F_{\mu\nu}(\rho)_{ba}\overline{H}^{\,(Q)}_a\rangle+h.c.\Big]
\end{eqnarray}
where
$F_{\mu\nu}(\rho)=\partial_{\mu}\rho_{\nu}-\partial_{\nu}\rho_{\mu}+[\rho_{\mu},\rho_{\nu}]$,
and $\rho_{\mu}$ is defined as
\begin{equation}
\label{a2}\rho_{\mu}=i\frac{g_V}{\sqrt{2}}V_{\mu}
\end{equation}
$V_{\mu}$ is a hermitian $3\times3$ matrix analogous to Eq.(\ref{5})
containing $\rho$, $K^{*}$, $\omega$ and $\phi$,
\begin{equation}
\label{a3}V=\left(\begin{array}{ccc}
\frac{\rho^0}{\sqrt{2}}+\frac{\omega}{\sqrt{2}}&\rho^{+}&K^{*+}\\
\rho^{-}&-\frac{\rho^0}{\sqrt{2}}+\frac{\omega}{\sqrt{2}}&K^{*0}\\
K^{*-}&\overline{K}^{*0}&\phi
\end{array}
\right)
\end{equation}
By imposing the KSRF relations, one obtains $g_V\simeq5.8$. For the
same reason, the interaction between the light vector resonances and
heavy anti-mesons should be included via applying charge conjugation
transformation
\begin{eqnarray}
\nonumber&&{\cal L}'_{V}=-i\beta\langle
\overline{H}^{\,(\overline{Q})}_av^{\mu}({\cal
V}_{\mu}-\rho_{\mu})_{ab}H^{(\overline{Q})}_b\rangle+i\lambda\langle
\overline{H}^{\,(\overline{Q})}_a\sigma^{\mu\nu}F_{\mu\nu}(\rho)_{ab}H^{(\overline{Q})}_b\rangle-i\beta_1\langle\overline{S}^{\,(\overline{Q})}_av^{\mu}({\cal V}_{\mu}-\rho_{\mu})_{ab}S^{\,(\overline{Q})}_b\rangle\\
\nonumber&&+i\lambda_1\langle\overline{S}^{\,(\overline{Q})}_a\sigma^{\mu\nu}F_{\mu\nu}(\rho)_{ab}S^{(\overline{Q})}_b\rangle-i\beta_2\langle\overline{T}^{\,(\overline{Q})}_{a\lambda}v^{\mu}({\cal V}_{\mu}-\rho_{\mu})_{ab}T^{(\overline{Q})\lambda}_{b}\rangle+i\lambda_2\langle\overline{T}^{\,(\overline{Q})}_{a\lambda}\sigma^{\mu\nu}F_{\mu\nu}(\rho)_{ab}T^{(\overline{Q})\lambda}_{b}\rangle\\
\nonumber&&+\Big[i\zeta\langle\overline{S}^{\,(\overline{Q})}_a\gamma^{\mu}({\cal V}_{\mu}-\rho_{\mu})_{ab}H^{(\overline{Q})}_a\rangle+i\mu\langle\overline{S}^{\,(\overline{Q})}_a\sigma^{\lambda\nu}F_{\lambda\nu}(\rho)_{ab}H^{(\overline{Q})}_b\rangle-i\zeta_1\langle\overline{H}^{\,(\overline{Q})}_a({\cal V}_{\mu}-\rho_{\mu})_{ab}T^{(\overline{Q})\mu}_b\rangle\\
\label{a4}&&+\mu_1\langle\overline{H}^{\,(\overline{Q})}_a\gamma^{\nu}F_{\mu\nu}(\rho)_{ab}T^{(\overline{Q})\mu}_b\rangle+h.c.\Big]
\end{eqnarray}
where we have used the property ${\cal C}V_{\mu}{\cal
C}^{-1}=-V^{T}_{\mu}$. Then the effective interactions relevant to
the concerned tree level scattering diagrams are as follows
\begin{eqnarray}
\nonumber&&{\cal L}_{{\rm DD}V}= ig_{{\rm DD}V}({\rm
D}_b\stackrel{\leftrightarrow}{\partial_{\mu}}{\rm
D}^{\dagger}_a)V^{\mu}_{ba}+ig_{\overline{\rm D}\,\overline{\rm
D}V}(\overline{\rm
D}_b\stackrel{\leftrightarrow}{\partial_{\mu}}\overline{\rm
D}^{\dagger}_a)V^{\mu}_{ab}\\
\nonumber&&{\cal L}_{{\rm D_1D_1}V}= ig_{{\rm D_1D_1}V}({\rm
D}^{\nu}_{1b}\stackrel{\leftrightarrow}{\partial_{\mu}}{\rm
D}^{\dagger}_{1a\nu})V^{\mu}_{ba}+ig'_{{\rm D_1D_1}V}({\rm
D}^{\mu}_{1b}{\rm D}^{\nu\dagger}_{1a}-{\rm D}^{\mu\dagger}_{1a}{\rm
D}^{\nu}_{1b})(\partial_{\mu}V_{\nu}-\partial_{\nu}V_{\mu})_{ba}\\
\nonumber&&+ig_{\overline{\rm D}_1\overline{\rm D}_1V}(\overline{\rm D}_{1b\nu}\stackrel{\leftrightarrow}{\partial_{\mu}}\overline{\rm D}^{\nu\dagger}_{1a})V^{\mu}_{ab}+ig'_{\overline{\rm D}_1\overline{\rm D}_1V}(\overline{\rm D}^{\mu}_{1b}\overline{\rm D}^{\nu\dagger}_{1a}-\overline{\rm D}^{\mu\dagger}_{1a}\overline{\rm D}^{\nu}_{1b})(\partial_{\mu}V_{\nu}-\partial_{\nu}V_{\mu})_{ab}\\
\nonumber&&{\cal L}_{{\rm DD_1}V}=g_{{\rm DD_1}V}{\rm
D}^{\mu}_{1b}V_{\mu ba}{\rm D}^{\dagger}_a+g'_{{\rm DD_1}V}({\rm
D}^{\mu}_{1b}\stackrel{\leftrightarrow}{\partial^{\nu}}{\rm
D}^{\dagger}_a)(\partial_{\mu}V_{\nu}-\partial_{\nu}V_{\mu})_{ba}\\
\nonumber&&+g_{\overline{\rm D}\,\overline{\rm D}_1 V}\overline{\rm
D}^{\dagger}_{a}V_{\mu ab}\overline{\rm
D}^{\mu}_{1b}+g'_{\overline{\rm D}\,\overline{\rm
D}_1V}(\overline{\rm
D}^{\mu}_{1b}\stackrel{\leftrightarrow}{\partial^{\nu}}\overline{\rm
D}^{\dagger}_a)(\partial_{\mu}V_{\nu}-\partial_{\nu}V_{\mu})_{ab}+h.c.\\
\nonumber&&{\cal L}_{{\rm D^{*}D^{*}}V}= ig_{{\rm D^{*}D^{*}}V}({\rm
D}^{*}_{b\nu}\stackrel{\leftrightarrow}{\partial_{\mu}}{\rm
D}^{*\nu\dagger}_a)V^{\mu}_{ba}+ig'_{{\rm D^{*}D^{*}}V}({\rm
D}^{*\mu}_{b}{\rm D}^{*\nu\dagger}_{a}-{\rm D}^{*\mu\dagger}_{a}{\rm
D}^{*\nu}_{b})(\partial_{\mu}V_{\nu}-\partial_{\nu}V_{\mu})_{ba}\\
\nonumber&&+ ig_{\overline{\rm D}^{*}\overline{\rm
D}^{*}V}(\overline{\rm
D}^{*}_{b\nu}\stackrel{\leftrightarrow}{\partial_{\mu}}\overline{\rm D}^{*\nu\dagger}_{a})V^{\mu}_{ab}+ig'_{\overline{\rm D}^{*}\overline{\rm D}^{*}V}(\overline{\rm D}^{*\mu}_b\overline{\rm D}^{*\nu\dagger}_a-\overline{\rm D}^{*\mu\dagger}_a\overline{\rm D}^{*\nu}_b)(\partial_{\mu}V_{\nu}-\partial_{\nu}V_{\mu})_{ab}\\
\nonumber&&{\cal L}_{{\rm D_0D_0}V}=ig_{{\rm D_0D_0}V}({\rm
D}_{0b}\stackrel{\leftrightarrow}{\partial_{\mu}}{\rm
D}^{\dagger}_{0a})V^{\mu}_{ba}+ig_{\overline{\rm D}_0\overline{\rm
D}_0V}(\overline{\rm
D}_{0b}\stackrel{\leftrightarrow}{\partial_{\mu}}\overline{\rm
D}^{\dagger}_{0a})V^{\mu}_{ab}\\
\nonumber&&{\cal L}_{{\rm D^{*}D_0}V}= g_{{\rm D^{*}D_0}V}{\rm
D}^{*\mu}_bV_{\mu ba}{\rm D}^{\dagger}_{0a}+g'_{{\rm
D^{*}D_0}V}({\rm
D}^{*\nu}_{b}\stackrel{\leftrightarrow}{\partial^{\mu}}{\rm
D}^{\dagger}_{0a}-{\rm
D}^{*\mu}_b\stackrel{\leftrightarrow}{\partial^{\nu}}{\rm
D}^{\dagger}_{0a})(\partial_{\mu}V_{\nu}-\partial_{\nu}V_{\mu})_{ba}\\
\label{a5}&&+g_{\overline{\rm D}^{*}\overline{\rm D}_0
V}\overline{\rm D}^{\dagger}_{0a}V_{\mu ab}\overline{\rm
D}^{*\mu}_{b}+g'_{\overline{\rm D}^{*}\overline{\rm
D}_0V}(\overline{\rm
D}^{*\nu}_b\stackrel{\leftrightarrow}{\partial^{\mu}}\overline{\rm
D}^{\dagger}_{0a}-\overline{\rm
D}^{*\mu}_b\stackrel{\leftrightarrow}{\partial^{\nu}}\overline{\rm
D}^{\dagger}_{0a})(\partial_{\mu}V_{\nu}-\partial_{\nu}V_{\mu})_{ab}+h.c.
\end{eqnarray}
The coupling constants are as follows\footnote[1]{The following
effective Lagrangian are obtained by expanding Eq.(\ref{a1}) and
Eq.(\ref{a4}) term by term, then they are checked by the program
FeynCalc\cite{feyncalc}.}

\begin{eqnarray}
\nonumber&&g_{{\rm DD}V}=-g_{\overline{\rm D}\,\overline{\rm
D}V}=\frac{1}{\sqrt{2}}\beta g_{V}\\
\nonumber&&g_{{\rm D_1D_1}V}=-g_{{\overline{\rm D}_1\overline{\rm
D}_1}V}=\frac{1}{\sqrt{2}}\,\beta_2g_V\\
\nonumber&&g'_{{\rm D_1D_1}V}=-g'_{{\overline{\rm D}_1\overline{\rm
D}_1V}}=\frac{5\lambda_2g_V}{3\sqrt{2}}{\rm M_{D_1}}\\
\nonumber&&g_{{\rm DD_1}V}=-g_{{\overline{\rm D}\,\overline{\rm
D}_1}V}=-\frac{2}{\sqrt{3}}\,\zeta_1 g_V\sqrt{\rm M_{D}M_{D_1}}\\
\nonumber&&g'_{{\rm DD_1}V}=-g'_{{\overline{\rm D}\,\overline{\rm
D}_1}V}=\frac{1}{\sqrt{3}}\,\mu_1 g_V\\
\nonumber&&g_{{\rm D^{*}D^{*}}V}=-g_{{\overline{\rm
D}^{*}\overline{\rm D}^{*}}V}=-\frac{1}{\sqrt{2}}\,\beta g_{V}\\
\nonumber&&g'_{{\rm D^{*}D^{*}}V}=-g'_{{\overline{\rm
D}^{*}\overline{\rm D}^{*}}V}=-\sqrt{2}\;\lambda g_{V}{\rm
M_{D^{*}}}\\
\nonumber&&g_{{\rm D_0D_0}V}=-g_{\overline{\rm D}_0\overline{\rm
D}_0V}=-\frac{1}{\sqrt{2}}\,\beta_1g_V\\
\nonumber&&g_{{\rm D^{*}D_0}V}=g_{{\overline{\rm D}^{*}\overline{\rm
D}_0}V}=-\zeta g_V\sqrt{\rm 2M_{D^{*}}M_{D_0}}\\
\label{a6}&&g'_{{\rm D^{*}D_0}V}=g'_{\overline{\rm
D}^{*}\overline{\rm D}_0 V}=-\frac{1}{\sqrt{2}}\,\mu g_V
\end{eqnarray}
From the above effective interactions, following the general
procedure presented in section III, we can calculate the effective
potential from $\rho$ and $\omega$ exchange. For Y(4260), the
potential in coordinate space is
\begin{eqnarray}
\nonumber&&V^{\rho,\omega}_{12}(r)=V^{\rho,\omega}_{21}(r)=0\\
\nonumber&&V^{\rho,\omega}_{11}(r)=\frac{1}{4}\beta\beta_2g^2_V\Big[3H(\Lambda,m_{\rho},r)+H(\Lambda,m_{\omega},r)\Big]
+\frac{\beta\beta_2g^2_V({\rm M^2_D+M^2_{D_1}})}{\rm
32M^2_{D}M^2_{D_1}}\Big[3G(\Lambda,m_{\rho},r)+G(\Lambda,m_{\omega},r)\Big]\\
\nonumber&&-\frac{1}{6}g^2_V\Big[\zeta_1+\frac{\mu_1({\rm
M^2_D-M^2_{D_1}})}{2\sqrt{\rm
M_DM_{D_1}}}\Big]^2\Big[3Y(\Lambda,X_2,\mu_5,m_{\rho},r)+Y(\Lambda,X_2,\mu_6,m_{\omega},r)\Big]\\
\nonumber&&+\frac{g^2_V\mu^2_1({\rm M_D+M_{D_1}})^2}{\rm 72
M_DM_{D_1}}\Big[3Z(\Lambda,X_2,\mu_5,m_{\rho},r)+Z(\Lambda,X_2,\mu_6,m_{\omega},r)\Big]\\
\nonumber&&-\frac{g^2_V\zeta^2_1}{18}\Big[\frac{3}{m^2_{\rho}}Z(\Lambda,X_2,\mu_5,m_{\rho},r)+\frac{1}{m^2_{\omega}}Z(\Lambda,X_2,\mu_6,m_{\omega},r)\Big]\\
\nonumber&&V^{\rho,\omega}_{22}(r)=\frac{1}{4}\beta\beta_1g^2_V\Big[3H(\Lambda,m_{\rho},r)+H(\Lambda,m_{\omega},r)\Big]+\frac{\beta\beta_1g^2_V({\rm
M^2_{D^{*}}+M^2_{D_0}})}{\rm
32M^2_{D^*}M^2_{D_0}}\Big[3G(\Lambda,m_{\rho},r)+G(\Lambda,m_{\omega},r)\Big]\\
\nonumber&&-\frac{1}{4}g^2_V\Big[\zeta-\frac{\mu({\rm
M^2_{D^*}-M^2_{D_0}})}{\sqrt{\rm
M_{D^*}M_{D_0}}}\Big]^2\Big[3Y(\Lambda,X_3,\mu_7,m_{\rho},r)+Y(\Lambda,X_3,\mu_8,m_{\omega},r)\Big]\\
\nonumber&&+\frac{g^2_V\mu^2({\rm M_{D^*}+M_{D_0}})^2}{\rm
12M_{D^*}M_{D_0}}\Big[3Z(\Lambda,X_3,\mu_7,m_{\rho},r)+Z(\Lambda,X_3,\mu_8,m_{\omega},r)\Big]\\
\label{a7}&&-\frac{g^2_V\zeta^2}{12}\Big[\frac{3}{m^2_{\rho}}Z(\Lambda,X_3,\mu_7,m_{\rho},r)+\frac{1}{m^2_{\omega}}Z(\Lambda,X_3,\mu_8,m_{\omega},r)\Big]
\end{eqnarray}
The potential in coordinate space for ${\rm Z^{+}_2(4250)}$ is
\begin{eqnarray}
\nonumber&&V^{\rho,\omega}_{12}(r)=V^{\rho,\omega}_{21}(r)=0\\
\nonumber&&V^{\rho,\omega}_{11}(r)=-\frac{1}{4}\beta\beta_2g^2_V\Big[H(\Lambda,m_{\rho},r)-H(\Lambda,\omega,r)\Big]-\frac{\beta\beta_2g^2_V({\rm
M^2_D+M^2_{D1}})}{\rm
32M^2_{D}M^2_{D_1}}\Big[G(\Lambda,m_{\rho},r)-G(\Lambda,m_{\omega},r)\Big]\\
\nonumber&&-\frac{1}{6}g^2_V\Big[\zeta_1+\frac{\mu_1({\rm
M^2_D}-{\rm M^2_{D_1}})}{2\sqrt{\rm
M_DM_{D_1}}}\Big]^2\Big[Y(\Lambda,X_2,\mu_5,m_{\rho},r)-Y(\Lambda,X_2,\mu_6,m_{\omega},r)\Big]\\
\nonumber&&+\frac{g^2_V\mu^2_1({\rm M_D+M_{D_1}})^2}{\rm
72M_{D}M_{D_1}}\Big[Z(\Lambda,X_2,\mu_5,m_{\rho},r)-Z(\Lambda,X_2,\mu_6,m_{\omega},r)\Big]\\
\nonumber&&-\frac{g^2_V\zeta^2_1}{18}\Big[\frac{1}{m^2_{\rho}}Z(\Lambda,X_2,\mu_5,m_{\rho},r)-\frac{1}{m^2_{\omega}}Z(\Lambda,X_2,\mu_6,m_{\omega},r)\Big]\\
\nonumber&&V^{\rho,\omega}_{22}(r)=-\frac{1}{4}\beta\beta_1g^2_V\Big[H(\Lambda,m_{\rho},r)-H(\Lambda,m_{\omega},r)\Big]-\frac{\beta\beta_1g^2_V({\rm
M^2_{D^*}+M^2_{D_0}})}{\rm
32M^2_{D^*}M^2_{D_0}}\Big[G(\Lambda,m_{\rho},r)-G(\Lambda,m_{\omega},r)\Big]\\
\nonumber&&-\frac{1}{4}g^2_V\Big[\zeta-\frac{\mu({\rm
M^2_{D^{*}}-M^2_{D_0}})}{\sqrt{\rm
M_{D^{*}}M_{D_0}}}\Big]^2\Big[Y(\Lambda,X_3,\mu_7,m_{\rho},r)-Y(\Lambda,X_3,\mu_8,m_{\omega},r)\Big]\\
\nonumber&&+\frac{g^2_V\mu^2({\rm M_{D^{*}}+M_{D_0}})^2}{\rm
12M_{D^{*}}M_{D_0}}\Big[Z(\Lambda,X_3,\mu_7,m_{\rho},r)-Z(\Lambda,X_3,\mu_8,m_{\omega},r)\Big]\\
\label{a8}&&-\frac{g^2_V\zeta^2}{12}\Big[\frac{1}{m^2_{\rho}}Z(\Lambda,X_3,\mu_7,m_{\rho},r)-\frac{1}{m^2_{\omega}}Z(\Lambda,X_3,\mu_8,m_{\omega},r)\Big]
\end{eqnarray}
where the parameters $\mu_i(i=5,6,7,8)$ are given by
\begin{eqnarray}
\nonumber&&\mu^2_5=m^2_{\rho}-({\rm M_{D_1}-M_{D}})^2\\
\nonumber&&\mu^2_6=m^2_{\omega}-({\rm M_{D_1}-M_{D}})^2\\
\nonumber&&\mu^2_7=m^2_{\rho}-({\rm M_{D_0}-M_{D^{*}}})^2\\
\label{a9}&&\mu^2_8=m^2_{\omega}-({\rm M_{D_0}-M_{D^{*}}})^2
\end{eqnarray}
The new functions $G(\Lambda,m,r)$ and $Y(\Lambda,X,\mu,m,r)$ are
defined as follows
\begin{eqnarray}
\nonumber&&G(\Lambda,m,r)=\frac{1}{4\pi}\frac{1}{r}(\Lambda^2e^{-\Lambda
r}-m^2e^{-mr})+\frac{\Lambda^2-m^2}{8\pi}(\Lambda-\frac{2}{r})e^{-\Lambda
r}\\
\label{a10}&&Y(\Lambda,X,\mu,m,r)=\frac{1}{4\pi}\frac{1}{r}(e^{-\mu
r}-e^{-Xr})-\frac{\Lambda^2-m^2}{8\pi X}e^{-Xr}
\end{eqnarray}
As is demonstrated in Eq.(\ref{a7}) and Eq.(\ref{a8}), light vector
mesons $\rho$ and $\omega$ exchange leads to diagonal interaction,
and the off-diagonal components of the effective potential are zero
because of parity conservation. For ${\rm Z^{+}_2(4250)}$, it is
obvious that the potential coming from $\rho$ exchange almost
cancels that from $\omega$ exchange, because of the accidental
coincidence of $m_{\rho}$ and $m_{\omega}$, i.e.,
$m_{\rho}\simeq775.5$ MeV and $m_{\omega}\simeq782.7$ MeV\cite{pdg}.

There are a number of parameters $\beta$, $\beta_1$, $\beta_2$,
$\mu$, $\mu_1$, $\zeta$ and $\zeta_1$ involved in the effective
potential. The information about the effective coupling constants
between the heavy meson and the light vector mesons is very scarce
until now, especially those related with the P-wave heavy mesons. By
vector meson dominance, $\beta$ is estimated to be about 0.9
\cite{Isola:2003fh}. Ref.\cite{Casalbuoni:1996pg} gives
$\mu=-0.1{\rm GeV^{-1}}$ and $\zeta=0.1$. The remaining parameters
have not been determined as far as we know, and we even don't know
the ranges which they are in. So at present we can not give a
quantitative estimate about the vector meson exchange contribution
to the potential associated with Y(4260). Since the light vector
meson mass $m_{\rho}$, $m_{\omega}$ is larger than $m_{\pi}$,
$m_{\eta}$ and $m_{\sigma}$, we expect that the potential induced by
vector meson exchange should be smaller than that due to
pseudoscalar and scalar exchange in magnitude. In principle, we can
determine these coupling constants following the methods of QCD sum
rule, non-relativistic potential model and so on, by means of which
certain coupling constants in HM$\chi$PT have been estimated. In
future, if we could get a reliable estimate about these coupling
constants from both phenomenological and theoretical approaches, The
effective potential arising from $\rho$, $\omega$ exchange and its
effect on the structure of Y(4260) could be analyzed in the same way
as in section V.
\end{appendix}

\newpage

\begin{center}
\begin{table}
\begin{tabular}{|c|cccc||cccc|}\hline\hline
\multicolumn{1}{|c}{}&\multicolumn{4}{|c||}{}&\multicolumn{4}{c|}{}\\
\multicolumn{1}{|c}{}&\multicolumn{4}{|c||}{${\rm Y_{cc}}{\rm ~with~
I^{G}(J^{PC})=0^{-}(1^{--})}$}&\multicolumn{4}{c|}{${\rm
Z^{+}_{cc}}{\rm ~with~ I^{G}(J^{P})=1^{-}(1^{-})}$}\\\hline
$g\tilde{h}$&$\Lambda({\rm GeV})$&${\rm M}(\rm MeV)$&${\rm r}_{\rm
rms}({\rm fm})$&{\rm R}&$\Lambda({\rm GeV})$&${\rm M}({\rm
MeV})$&${\rm r}_{\rm rms}({\rm fm})$&{\rm R}\\\hline

    &3.3&-4.04&1.39&2.63&14.5&-2.40&1.76&3.27\\
0.23    &3.4&-12.20&0.82&1.71&14.6&-7.53&0.98&2.02\\
    &3.5&-23.86&0.63&1.41&14.7&-14.71&0.71&1.60\\\hline

    &2.3&-3.29&1.56&2.90&9.6&-3.53&1.45&2.76\\
0.35    &2.4&-11.32&0.86&1.74&9.7&-9.24&0.89&1.87\\
    &2.5&-24.95&0.63&1.40&9.8&-16.99&0.68&1.54\\\hline

    &1.6&-1.79&2.17&3.88&6.3&-5.53&1.17&2.29\\
0.54&1.7&-10.33&0.95&1.84&6.4&-12.13&0.80&1.71\\
    &1.8&-24.66&0.67&1.41&6.5&-20.82&0.63&1.45\\\hline

    &1.2&-7.43&1.15&2.11&4.0&-3.45&1.50&2.81\\
0.85&1.3&-22.69&0.76&1.46&4.1&-9.28&0.93&1.88\\
    &1.4&-46.18&0.57&1.24&4.2&-17.42&0.71&1.53\\\hline\hline
\end{tabular}
\caption{\label{charm-table1}The mass, the root of mean square
radius(rms) and the ratio(R) between the ${\rm DD_1}$ and ${\rm
D^{*}D_0}$ components for the bound state solutions of the ${\rm
DD_1}$ and ${\rm D^{*}D_0}$ system with one pseudoscalar exchange,
and the mass is measured with respect to the ${\rm D_1D}$ threshold
${\rm M_{D}+M_{D_1}\simeq4291.3MeV}$.}
\end{table}
\end{center}
\newpage

\begin{center}
\begin{table}[hptb]
\begin{tabular}{|c|c|c|c|cccc||cccc|}\hline\hline
\multicolumn{4}{|c}{}&\multicolumn{4}{|c||}{}&\multicolumn{4}{c|}{}\\
\multicolumn{4}{|c}{}&\multicolumn{4}{|c||}{${\rm Y_{cc}}{\rm ~with~
I^{G}(J^{PC})=0^{-}(1^{--})}$}&\multicolumn{4}{c|}{${\rm
Z^{+}_{cc}}{\rm ~with~ I^{G}(J^{P})=1^{-}(1^{-})}$}\\\hline
$g\tilde{h}$&$g_{\sigma}.g'_{\sigma}$&$g_{\sigma}.g''_{\sigma}$&$h_{\sigma}$&$\Lambda({\rm
GeV})$&${\rm M}(\rm MeV)$&${\rm r}_{\rm rms}({\rm fm})$&{\rm
R}&$\Lambda({\rm GeV})$&${\rm M}({\rm MeV})$&${\rm r}_{\rm rms}({\rm
fm})$&{\rm R}\\\hline

    &    &    &    &4.7&-3.43&1.48&2.81&6.1&-5.43&1.09&1.47\\
    &    &    & 0.35&4.8&-10.87&0.84&1.81&6.2&-15.82&0.66&1.00\\
    &0.58&0.58&    &4.9&-21.42&0.62&1.50&6.3&-29.69&0.51&0.84\\\cline{4-12}
    &    &    &    &8.7&-5.49 & 1.16&2.43&4.0&-5.49&1.06&1.19\\
    &    &    &0.50&8.8&-12.52&0.78 &1.82 &4.1&-18.45&0.62&0.74\\
    &    &    &    &8.9&-21.75 &0.61 & 1.58&4.2&-36.23&0.48&0.59\\\cline{2-12}

    &    &    &   &4.5&-3.65&1.50&3.49  &5.9&-4.22&1.33&2.52\\
    &    &    &0.35&4.6& -9.70&0.93 &2.31 &6.0&-11.84&0.80&1.59\\
    &0.58&-0.58&    &4.7& -18.27&0.70 &1.86 &6.1&-22.48&0.60&1.25\\\cline{4-12}
    &    &    &    &8.4&-8.62&0.98 &2.62 &3.9&-5.62&1.14&1.94\\
    &    &    &0.50&8.5&-15.03 & 0.75 &2.14&4.0&-16.19&0.70&1.17\\
    &    &    &     &8.6&-23.09 &0.62 &1.88&4.1&-31.14&0.53&0.90\\\cline{2-12}

0.23&    &   &   &4.6&-7.02 &1.02 &1.80 &5.9&-8.06&0.87&0.95\\
    &    &    &0.35&4.7& -15.40& 0.72 &1.41&6.0&-18.32&0.61&0.73\\
    &-0.58&0.58&    &4.8& -26.34 & 0.58&1.25&6.1&-31.34&0.50&0.64\\\cline{4-12}
    &    &    &     &8.4&-3.81&1.37 &2.32  &3.9&-11.21&0.74&0.65\\
    &    &    &0.50 &8.5& -9.37 &0.88 &1.67 &4.0&-25.48&0.54&0.50\\
    &    &    &     &8.6&-16.74 & 0.68 &1.42&4.1&-43.76&0.45&0.43\\\cline{2-12}

    &    &    &    &4.4&-6.47 &1.11 &2.32&5.7&-5.35&1.15&1.69\\
    &    &    &0.35&4.5& -13.31&0.80& 1.78&5.8&-12.98&0.75&1.17\\
    &-0.58&-0.58&   &4.6&-22.33&0.64&1.53 &5.9&-23.10&0.59&0.96\\\cline{4-12}
    &    &    &     &8.0&-3.13 & 1.60 &3.37&3.8&-8.99&0.88&1.12\\
    &    &    &0.50 &8.1& -7.05& 1.06 &2.36 &3.9&-20.74&0.62&0.78\\
    &    &    &     &8.2& -12.35 &0.82 &1.92 &4.0&-36.29&0.50&0.64\\\hline

    &    &    &   &2.8&-2.15 &1.91 &3.38 &5.2&-8.56&0.90&1.48\\
    &    &    &0.35&2.9& -9.48&0.93&1.85 &5.3&-19.71&0.62&1.13\\
    &0.58&0.58&    &3.0&-20.76&0.66& 1.46 &5.4&-34.24&0.50&0.98\\\cline{4-12}
    &    &    &     &3.7&-6.18&1.12&2.19 &3.6&-3.77&1.34&1.87\\
    &    &    &0.50 &3.8&-14.57 &0.76 &1.64&3.7&-14.71&0.72&1.07\\
    &    &    &     &3.9&-25.91 &0.59 &1.42&3.8&-30.46&0.53&0.85\\\cline{2-12}

    &    &    &  &2.8&-9.66&0.95&2.15&5.0&-6.18&1.12&2.38\\
    &    &    &0.35&2.9& -19.82 &0.70&1.70&5.1&-14.42&0.76&1.68\\
    &0.58&-0.58&    &3.0&-33.45 &0.56&1.48&5.2&-25.58&0.58&1.38\\\cline{4-12}
    &    &    &     &3.5&-3.96 &1.45&3.20 &3.5&-4.65&1.29&2.55\\
    &    &    &0.50 &3.6&-10.06 &0.93&2.18  &3.6&-13.97&0.77&1.54\\
    &    &    &     &3.7&-18.66&0.70 &1.77 &3.7&-27.47&0.57&1.18\\\cline{2-12}

0.35&    & &  &2.8&-8.07 &0.99 &1.74 &5.0&-7.24&0.95&1.25\\
    &    &    &0.35&2.9& -18.21&0.70&1.36 &5.1&-16.94&0.66&0.94\\
    &-0.58&0.58&    &3.0&-31.84 &0.56 &1.21 &5.2&-29.51&0.53&0.82\\\cline{4-12}
    &    &    &    &3.6&-7.83 & 0.99&1.77&3.5&-5.98&1.03&1.19\\
    &    &    & 0.50&3.7&-16.23& 0.72&1.42  &3.6&-17.51&0.65&0.80\\
    &    &    &     &3.8&-27.23 &0.59 &1.26 &3.7&-33.14&0.51&0.66\\\cline{2-12}

    &    &   &    &2.7&-8.08 &1.03 &2.05 &4.9&-11.84&0.81&1.44\\
    &    &    &0.35   &2.8&-17.20 &0.74 &1.59 &5.0&-21.48&0.63&1.17\\
    &-0.58 &-0.58&       &2.9&-29.48&0.59 &1.38 &5.1&-33.64&0.52&1.03\\\cline{4-12}
    &    &    &       &3.4&-5.08 &1.27&2.51 &3.4&-5.68&1.14&1.80\\
    &    &    &0.50   &3.5&-11.34 & 0.87 &1.84&3.5&-15.32&0.72&1.16\\
    &    &    &       &3.6&-19.83&0.69 &1.55&3.6&-28.71&0.56&0.93\\\hline\hline

\end{tabular}
\caption{\label{charm-table2}The mass, the root of mean square
radius(rms) and the ratio(R) between the ${\rm DD_1}$ and ${\rm
D^{*}D_0}$ components for the bound state solutions of the ${\rm
DD_1}$ and ${\rm D^{*}D_0}$ system with both one pseudoscalar
exchange and $\sigma$ exchange, and the mass is measured with
respect to the ${\rm D_1D}$ threshold ${\rm
M_{D}+M_{D_1}\simeq4291.3MeV}$.}
\end{table}
\end{center}

\newpage

\begin{center}
\begin{table}[hptb]
\begin{tabular}{|c|c|c|c|cccc||cccc|}\hline\hline
\multicolumn{4}{|c}{}&\multicolumn{4}{|c||}{}&\multicolumn{4}{c|}{}\\
\multicolumn{4}{|c}{}&\multicolumn{4}{|c||}{${\rm Y_{cc}}{\rm ~with~
I^{G}(J^{PC})=0^{-}(1^{--})} $}&\multicolumn{4}{c|}{${\rm
Z^{+}_{cc}}{\rm ~with~ I^{G}(J^{P})=1^{-}(1^{-})}$}\\\hline
$g\tilde{h}$&$g_{\sigma}.g'_{\sigma}$&$g_{\sigma}.g''_{\sigma}$&$h_{\sigma}$&$\Lambda({\rm
GeV})$&${\rm M}(\rm MeV)$&${\rm r}_{\rm rms}({\rm fm})$&{\rm
R}&$\Lambda({\rm GeV})$&${\rm M}({\rm MeV})$&${\rm r}_{\rm rms}({\rm
fm})$&{\rm R}\\\hline

    &    &   &    &1.8&-2.42&1.84&3.21&4.2&-9.99&0.86&1.59\\
    &    &    &0.35 &1.9&-11.03&0.91&1.75&4.3&-21.32&0.62&1.25\\
    &0.58&0.58&     &2.0&-24.76&0.66&1.39&4.4&-36.09&0.50&1.10\\\cline{4-12}
    &    &    &     &2.1&-8.24 &1.02&1.92&3.2&-12.62&0.78&1.38\\
    &    &    &0.50 &2.2&-19.25 &0.71&1.47 &3.3&-26.99&0.57&1.08\\
    &    &    &     &2.3&-34.54 &0.57&1.30 &3.4&-45.95&0.47&0.95\\\cline{2-12}

    &    &    &    &1.8&-6.29 &1.19&2.42&4.1&-13.88&0.78&1.83\\
    &    &    &0.35&1.9&-16.88&0.78&1.69&4.2&-24.83&0.61&1.50\\
    &0.58&-0.58&    &2.0&-32.31&0.60&1.43&4.3&-38.79&0.50&1.32\\\cline{4-12}
    &    &    &    &2.0&-5.65 &1.25&2.51 &3.1&-11.78&0.85&1.87\\
    &    &    &0.50&2.1&-14.59&0.82&1.77 &3.2&-24.09&0.62&1.43\\
    &    &    &     &2.2&-27.44&0.64&1.49&3.3&-40.53&0.51&1.21\\\cline{2-12}

0.54&    &   &    &1.8&-5.35&1.25&2.11&4.1&-13.94&0.74&1.18\\
    &    &    &0.35  &1.9&-15.93&0.78&1.44 &4.2&-25.68&0.58&1.00\\
    &-0.58&0.58&      &2.0&-31.41 &0.60&1.23&4.3&-40.44&0.48&0.91\\\cline{4-12}
    &    &    &      &2.0&-4.61 &1.33&2.20 &3.1&-12.19&0.79&1.16\\
    &    &    &0.50  &2.1&-13.49&0.83&1.50 &3.2&-25.70&0.59&0.92\\
    &    &    &      &2.2&-26.33 &0.64&1.27&3.3&-43.37&0.48&0.81\\\cline{2-12}

    &    &   &    &1.8&-9.78&0.98 &1.88  &3.9&-8.33&0.98&1.84\\
    &    &    & 0.35  &1.9&-22.04 &0.70&1.46&4.0&-16.85&0.71&1.42\\
    &-0.58&-0.58&       &2.0&-39.09&0.56&1.28&4.1&-28.01&0.58&1.22\\\cline{4-12}
    &    &    &       &2.0&-9.66 &0.98&1.87&3.0&-10.72&0.87&1.60\\
    &    &    & 0.50  &2.1&-20.31&0.72&1.48&3.1&-22.23&0.64&1.22\\
    &    &    &       &2.2&-34.80&0.58&1.31&3.2&-37.53&0.52&1.04\\\hline

    &    &    &    &1.3&-11.86&0.93&1.72&3.2&-10.84&0.86&1.67\\
    &    &    &0.35 &1.4&-28.73&0.67&1.34&3.3&-22.29&0.64&1.34\\
    &0.58&0.58&     &1.5&-53.33 &0.53&1.20 &3.4&-37.36&0.52&1.19\\\cline{4-12}
    &    &    &     &1.3&-5.27 & 1.31&2.29 &2.6&-9.21&0.94&1.76\\
    &    &    &0.50 &1.4&-16.97&0.81&1.52 &2.7&-21.74&0.65&1.32\\
    &    &    &     &1.5&-34.96&0.62&1.28&2.8&-38.91&0.52&1.13\\\cline{2-12}

    &    &    &    &1.3&-14.66&0.87&1.71&3.1&-11.85&0.86&1.99\\
    &    &    & 0.35 &1.4&-32.74&0.64&1.38&3.2&-22.30&0.66&1.61\\
    &0.58&-0.58&      &1.5&-58.63&0.52&1.24 &3.3&-35.98&0.54&1.41\\\cline{4-12}
    &    &    &      &1.3&-7.58 &1.14&2.13&2.5&-7.56&1.07&2.38\\
    &    &    & 0.50 &1.4&-20.52&0.76&1.53 &2.6&-17.94&0.73&1.70\\
    &    &    &      &1.5&-39.78&0.60 &1.31&2.7&-32.49&0.57&1.40\\\cline{2-12}

0.85&    &   &    &1.3&-14.13&0.87&1.56&3.1&-10.85&0.85&1.44\\
    &    &    &0.35  &1.4&-32.23&0.64&1.26&3.2&-21.69&0.64&1.17\\
    &-0.58&0.58&      &1.5&-58.16&0.52&1.13&3.3&-35.80&0.53&1.05\\\cline{4-12}
    &    &    &      &1.3&-7.04 &1.15&1.95 &2.5&-6.58&1.08&1.74\\
    &    &    & 0.50 &1.4&-19.98&0.76&1.38&2.6&-17.47&0.71&1.23\\
    &    &    &      &1.5&-39.26 &0.59&1.20&2.7&-32.64&0.56&1.03\\\cline{2-12}

    &    &   &    &1.2&-4.75&1.40 &2.51&3.0&-11.33&0.87&1.75\\
    &    &    & 0.35 &1.3&-16.97&0.82&1.56&3.1&-21.18&0.67&1.42\\
    &-0.58&-0.58&      &1.4&-36.26&0.62&1.29&3.2&-33.99&0.56&1.25\\\cline{4-12}
    &    &    &      &1.3&-9.47 &1.03&1.87 &2.5&-13.90&0.81&1.62\\
    &    &    &0.50  &1.4&-23.58&0.72&1.41&2.6&-26.68&0.62&1.30\\
    &    &    &      &1.5&-44.09&0.58&1.23&2.7&-43.54&0.51&1.14\\\hline\hline

\end{tabular}
\caption{\label{charm-table3}The continuing of Table
\ref{charm-table2}.}
\end{table}
\end{center}

\newpage

\begin{center}
\begin{table}[hptb]
\begin{tabular}{|c|cccc||cccc|}\hline\hline
\multicolumn{1}{|c}{}&\multicolumn{4}{|c||}{}&\multicolumn{4}{c|}{}\\
\multicolumn{1}{|c}{}&\multicolumn{4}{|c||}{${\rm Y_{bb}}{\rm ~with~
I^{G}(J^{PC})=0^{-}(1^{--})} $}&\multicolumn{4}{c|}{${\rm
Z^{+}_{bb}}{\rm ~with~ I^{G}(J^{P})=1^{-}(1^{-})}$}\\\hline
$g\tilde{h}$&$\Lambda({\rm GeV})$&${\rm M}(\rm MeV)$&${\rm r}_{\rm
rms}({\rm fm})$&{\rm R}&$\Lambda({\rm GeV})$&${\rm M}({\rm
MeV})$&${\rm r}_{\rm rms}({\rm fm})$&{\rm R}\\\hline

    & 1.8&-5.60&0.81 &3.08 &6.2 &-7.90&0.63&2.53 \\
0.23& 1.9&-14.96&0.53 &2.03 &6.3 &-13.45 &0.49 &2.04 \\
    & 2.0&-28.27&0.42&1.63 & 6.4&-20.22 & 0.40&1.76 \\\hline

    & 1.4&-8.76 &0.70 &2.57 & 4.2&-5.17 & 0.88&3.08 \\
0.35& 1.5&-21.93 &0.49 &1.79 &4.3 &-10.31 &0.57 &2.29 \\
    & 1.6& -40.65&0.39 &1.48 &4.4 &-16.95 &0.45 &1.89\\\hline

    &1.1 &-11.29 &0.66 &2.35 &2.9 &-6.55& 0.73&2.81 \\
0.54&1.2 &-29.05 & 0.47&1.65 & 3.0& -12.79&0.54 & 2.13\\
    & 1.3&-54.93&0.38 &1.38 &3.1 &-20.87 & 0.44&1.77 \\\hline

    & 0.8&-3.49 & 1.13&4.14 &2.0 &-7.38 &0.72 &2.72 \\
0.85& 0.9& -18.98&0.58 &1.94 &2.1&-14.97& 0.53&2.03 \\
    & 1.0&-46.04 &0.43 &1.46&2.2 &-25.03 &0.44 &1.69 \\\hline\hline
 \end{tabular}
 \caption{\label{bottom-table1}The mass, the root of mean square
radius(rms) and the ratio(R) between the ${\rm BB_1}$ and ${\rm
B^{*}B_0}$ components for the bound state solutions of the ${\rm
BB_1}$ and ${\rm B^{*}B_0}$ system with one pseudoscalar exchange,
and the mass is measured with respect to the ${\rm B_1B}$ threshold
${\rm M_{B}+M_{B_1}\simeq11004MeV.}$.}
\end{table}
\end{center}

\newpage

\begin{center}
\begin{table}[hptb]
\begin{tabular}{|c|c|c|c|cccc||cccc|}\hline\hline
\multicolumn{4}{|c}{}&\multicolumn{4}{|c||}{}&\multicolumn{4}{c|}{}\\
\multicolumn{4}{|c}{}&\multicolumn{4}{|c||}{${\rm Y_{bb}}{\rm ~with~
I^{G}(J^{PC})=0^{-}(1^{--})}$}&\multicolumn{4}{c|}{${\rm
Z^{+}_{bb}}{\rm ~with~ I^{G}(J^{P})=1^{-}(1^{-})}$}\\\hline
$g\tilde{h}$&$g_{\sigma}.g'_{\sigma}$&$g_{\sigma}.g''_{\sigma}$&$h_{\sigma}$&$\Lambda({\rm
GeV})$&${\rm M}(\rm MeV)$&${\rm r}_{\rm rms}({\rm fm})$&{\rm
R}&$\Lambda({\rm GeV})$&${\rm M}({\rm MeV})$&${\rm r}_{\rm rms}({\rm
fm})$&{\rm R}\\\hline

    &    &    &    & 2.4&-8.16&0.65 &2.47 &3.1 &-4.84&0.76&2.21\\
    &    &    &0.35&2.5 &-17.22 &0.47 &1.89 &3.2 &-14.28&0.59&1.41 \\
    &0.58&0.58&    &2.6 &-29.14 & 0.39&1.63 &3.3 &-27.06&0.36&1.11 \\\cline{4-12}
    &    &    &    &3.8 &-4.40 & 0.87 &3.44 &2.3 &-11.53&0.52&1.42\\
    &    &    &0.50& 3.9&-9.54 &0.59  &2.50  & 2.4&-27.04 &0.38&0.94 \\
    &    &    &    & 4.0&-16.34  &0.46  &2.07  &2.5&-47.66&0.31&0.74\\\cline{2-12}

    &    &    &   & 2.2&-5.63 &0.84 &3.87   & 2.9&-5.32&0.83&4.34 \\
    &    &    &0.35& 2.3&-12.07  &0.59  & 2.73 &3.0 &-11.82&0.57&2.78\\
    &0.58&-0.58&    &2.4 &-20.93  &0.47  &2.19  &3.1 &-20.94&0.44&2.05 \\\cline{4-12}
    &    &    &    &3.5 &-7.67 & 0.73 &3.76  &2.2 &-11.16&0.60&2.84\\
    &    &    &0.50&3.6 & -12.17 & 0.58  &3.02 &2.3 &-23.19&0.44&1.78\\
    &    &    &     & 3.7& -17.83 &0.49  &2.57&2.4 &-39.83&0.35&1.29\\\cline{2-12}

0.23&    &   &   & 2.3& -8.15 & 0.65 &2.17  &3.0 &-10.11&0.52&1.16 \\
    &    &    &0.35& 2.4& -16.87 & 0.48&1.69 & 3.1&-21.07&0.39&0.88 \\
    &-0.58&0.58&    & 2.5& -28.20 & 0.40 &1.46 &3.2 &-34.78&0.33&0.75 \\\cline{4-12}
    &    &    &    &3.6 &-5.80 & 0.75 &2.61 &2.2 &-10.04&0.54&1.08 \\
    &    &    &0.50&3.7 & -10.92&0.56&2.05  &2.3 &-24.96&0.38&0.71 \\
    &    &    &    & 3.8& -17.38 &0.45&1.76 &2.4 &-44.59&0.32&0.56 \\\cline{2-12}

    &    &   &  & 2.1& -5.02& 0.88&3.59 &2.8 &-7.04&0.71&2.64  \\
    &    &    &0.35& 2.2&-11.16 &0.61 & 2.48 &2.9 &-14.49&0.51&1.78 \\
    &-0.58&-0.58&    & 2.3&-19.60&0.48 &1.98  &3.0 &-24.44&0.41&1.38 \\\cline{4-12}
    &    &    &    & 3.3& -8.23 &0.70 &3.10 &2.1 &-8.20 &0.67&2.43 \\
    &    &    &0.50&3.4 &-12.63 &0.57&2.56& 2.2&-19.40&0.46&1.40 \\
    &    &    &     &3.5 &-18.04 &0.49 &2.21 &2.3 & -35.11&0.37&0.99 \\\hline

    &    &    &    &1.6 &-8.48&0.68&2.46 &2.7 &-8.38&0.61&2.01 \\
    &    &    &0.35&1.7 &-19.80&0.49&1.81 & 2.8&-19.05 &0.43&1.46 \\
    &0.58&0.58&    &1.8 &-35.44& 0.39&1.53 &2.9 &-33.04&0.35&1.21 \\\cline{4-12}
    &    &    &    &1.9 &-8.79& 0.66&2.43&2.1 &-11.60 &0.54 &1.72\\
    &    &    &0.50& 2.0&-17.95&0.49 &1.89 &2.2 &-26.56&0.40&1.19 \\
    &    &    &    &2.1 & -30.10 &0.41 & 1.63 &2.3 &-46.60&0.33&0.96\\\cline{2-12}

    &    &    &   &1.5 &-5.95&0.84&3.48 &2.5 &-6.02 &0.79&3.99 \\
    &    &    &0.35&1.6 &-15.06&0.57&2.32 &2.6 &-13.16 &0.55&2.65\\
    &0.58&-0.58&    &1.7 &-28.17 &0.45&1.85&2.7 &-23.07 & 0.44&2.02 \\\cline{4-12}
    &    &    &    &1.8 &-9.31&0.69&2.89&2.0 &-10.45 &0.62 &2.98\\
    &    &    & 0.50&1.9 &-17.68&0.52&2.23 & 2.1&-22.30&0.45&1.95\\
    &    &    &    &2.0 &-28.77&0.43&1.89& 2.2&-38.74&0.37 &1.47\\\cline{2-12}

0.35&    &   &   & 1.6&-12.71&0.58 &1.94&2.6 &-10.35 &0.55 &1.44 \\
    &    &    &0.35& 1.7&-25.80&0.44&1.54 &2.7 &-21.35 &0.41 &1.10 \\
    &-0.58&0.58&    & 1.8&-43.15&0.37&1.35 &2.8 &-35.32 &0.35&0.94 \\\cline{4-12}
    &    &    &    &1.8 &-6.49&0.76&2.51&2.0 &-8.08 &0.62&1.59 \\
    &    &    &0.50&1.9 &-14.61&0.54&1.85&2.1 &-21.54 &0.42&1.02 \\
    &    &    &    &2.0 &-25.57&0.43&1.56 & 2.2&-39.85 &0.35&0.80 \\\cline{2-12}

    &    &   &  &1.5 &-8.80&0.71&2.68 &2.4 &-6.03&0.78&3.09 \\
    &    &    &0.35&1.6 &-19.49&0.51&1.94&2.5 &-13.31 &0.54 &2.05 \\
    &-0.58&-0.58&    &1.7 &-34.19&0.42&1.61&2.6 &-23.26&0.43 &1.58 \\\cline{4-12}
    &    &    &    &1.7&-6.61&0.80&3.08&1.9 &-6.54 &0.76 &3.03 \\
    &    &    &0.50&1.8 &-13.96&0.58&2.23 &2.0 &-16.80 &0.50&1.77 \\
    &    &    &     &1.9&-23.92&0.47&1.84&2.1 &-31.57&0.40&1.27 \\\hline\hline

\end{tabular}
\caption{\label{bottom-table2}The mass, the root of mean square
radius(rms) and the ratio(R) between the ${\rm BB_1}$ and ${\rm
B^{*}B_0}$ components for the bound state solutions of the ${\rm
BB_1}$ and ${\rm B^{*}B_0}$ system with both one pseudoscalar
exchange and $\sigma$ exchange, and the mass is measured with
respect to the ${\rm B_1B}$ threshold ${\rm
M_{B}+M_{B_1}\simeq11004MeV}$.}
\end{table}
\end{center}

\newpage

\begin{center}
\begin{table}[hptb]
\begin{tabular}{|c|c|c|c|cccc||cccc|}\hline\hline
\multicolumn{4}{|c}{}&\multicolumn{4}{|c||}{}&\multicolumn{4}{c|}{}\\
\multicolumn{4}{|c}{}&\multicolumn{4}{|c||}{${\rm Y_{bb}}{\rm ~with~
I^{G}(J^{PC})=0^{-}(1^{--})}$}&\multicolumn{4}{c|}{${\rm
Z^{+}_{bb}}{\rm ~with~ I^{G}(J^{P})=1^{-}(1^{-})}$}\\\hline
$g\tilde{h}$&$g_{\sigma}.g'_{\sigma}$&$g_{\sigma}.g''_{\sigma}$&$h_{\sigma}$&$\Lambda({\rm
GeV})$&${\rm M}(\rm MeV)$&${\rm r}_{\rm rms}({\rm fm})$&{\rm
R}&$\Lambda({\rm GeV})$&${\rm M}({\rm MeV})$&${\rm r}_{\rm rms}({\rm
fm})$&{\rm R}\\\hline

    &    &    &    &1.1 &-4.16 &0.99 &3.49 &2.2 &-6.67&0.70 &2.45 \\
    &    &    &0.35&1.2 &-15.49&0.58 &2.00 &2.3 &-16.59 &0.48 &1.70 \\
    &0.58&0.58&    &1.3&-33.34&0.44&1.56 & 2.4&-30.12&0.38&1.39 \\\cline{4-12}
    &    &    &    &1.2 &-6.93&0.79&2.73 &1.8 &-6.41 &0.73 &2.56\\
    &    &    &0.50&1.3 &-18.44&0.53&1.87  &1.9 &-18.58 &0.47 &1.60 \\
    &    &    &    &1.4 &-35.15&0.43&1.55 & 2.0&-36.01 &0.37&1.24\\\cline{2-12}

    &    &    &   &1.1 &-6.49&0.84&3.19&2.1&-8.66 &0.68 &3.21\\
    &    &    &0.35&1.2 &-19.23&0.55&2.03&2.2 &-17.54&0.50&2.29\\
    &0.58&-0.58&    &1.3 &-38.42&0.43&1.62 &2.3 &-29.56 &0.41&1.83 \\\cline{4-12}
    &    &    &    &1.2 &-10.20& 0.69&2.60&1.7 &-5.05 &0.88 &4.32\\
    &    &    &0.50& 1.3&-23.12&0.51&1.90&1.8 &-14.36 &0.56 &2.48\\
    &    &    &     &1.4&-41.24&0.41&1.60&1.9 &-28.38 &0.43&1.79\\\cline{2-12}

0.54&    &   &   &1.1 &-5.38&0.88&2.98  &2.1 &-5.49 &0.76 &2.23 \\
    &    &    &0.35&1.2 & -17.94&0.55&1.81 &2.2 &-14.76 &0.50 &1.49 \\
    &-0.58&0.58&    &1.3 &-37.11&0.43&1.44& 2.3&-27.44&0.40&1.20 \\\cline{4-12}
    &    &    &    &1.2 &-8.84&0.71&2.35 &1.8 &-12.06 &0.56&1.62 \\
    &    &    &0.50&1.3 &-21.63&0.51&1.69 &1.9 &-27.02 &0.41 &1.16 \\
    &    &    &    &1.4 &-39.68&0.41&1.42&2.0 &-47.05 &0.34 &0.96 \\\cline{2-12}

    &    &   &  & 1.1&-7.79&0.78&2.81 &2.0 &6.63 &0.76 &3.06 \\
    &    &    &0.35& 1.2&-21.69&0.52&1.85&2.1 &-14.74 &0.54 &2.07 \\
    &-0.58&-0.58&    & 1.3&-42.15&0.42&1.50  &2.2 &-25.89&0.43&1.62 \\\cline{4-12}
    &    &    &    &1.2 &-12.19&0.65&2.30 & 1.7&-8.44&0.69&2.75 \\
    &    &    &0.50&1.3 &-26.32& 0.49 &1.73&1.8 &-20.06 &0.49&1.76 \\
    &    &    &    &1.4 &-45.74&0.40&1.49 &1.9 &-36.41 &0.39 &1.35 \\\hline

    &    &    &    & 0.9&-15.10 &0.63&2.09 &1.8 &-14.35 &0.54 &1.93 \\
    &    &    &0.35&1.0 &-37.85&0.46&1.53& 1.9&-27.75&0.42&1.53 \\
    &0.58&0.58&    &1.1 &-71.44 &0.37&1.31&2.0 &-45.31 & 0.35&1.32 \\\cline{4-12}
    &    &    &    &0.9 &-12.39&0.67&2.24 &1.5&-5.80&0.80 &2.90\\
    &    &    &0.50&1.0 &-31.80&0.48&1.60  &1.6 &-17.64 &0.51 &1.79 \\
    &    &    &    &1.1 &-60.44&0.39&1.36  &1.7 &-35.09&0.40 &1.39\\\cline{2-12}

    &    &    &   &0.9&-16.45 &0.62&2.10&1.7 &-11.83 &0.62&2.67 \\
    &    &    &0.35&1.0 &-40.19&0.45&1.56&1.8 &-23.08 &0.47 &2.00\\
    &0.58&-0.58&    &1.1& -74.91 &0.37&1.35&1.9 &-38.13 &0.39 &1.66 \\\cline{4-12}
    &    &    &    &0.9 & -13.70&0.65&2.25&1.5 &-11.33 &0.64 &2.73\\
    &    &    &0.50&1.0&-34.09 &0.48&1.64 &1.6 &-24.92 &0.47 &1.91\\
    &    &    &    &1.1 &-63.85&0.39&1.40&1.7&-43.85&0.38 &1.53\\\cline{2-12}

0.85&    &   &   &0.9 &-15.93&0.62&2.01&1.7 &-9.30 &0.64 &2.04 \\
    &    &    &0.35&1.0 &-39.56&0.45&1.47 &1.8 &-20.77 &0.47&1.49 \\
    &-0.58&0.58&    &1.1&-74.25&0.37&1.27&1.9 &-36.20&0.39&1.25 \\\cline{4-12}
    &    &    &    &0.9 &-13.17&0.66&2.15 &1.5 &-9.18 &0.65 &2.11 \\
    &    &    &0.50&1.0 &-33.43&0.48&1.54 &1.6 &-23.12&0.46&1.44 \\
    &    &    &    &1.1 &-63.13&0.39&1.31 &1.7 &-42.63 &0.38&1.16 \\\cline{2-12}

    &    &   &   &0.9 &-17.27&0.61&2.02 &1.6 &-7.05 &0.77&3.02 \\
    &    &    &0.35 &1.0 &-41.88&0.45&1.51 &1.7 &-16.42 &0.54 &2.04 \\
    &-0.58&-0.58&    &1.1 &-77.70&0.37&1.30&1.8 &-29.47&0.43 &1.61 \\\cline{4-12}
    &    &    &    &0.9 &-14.47&0.64&2.16 &1.5 &-14.85 &0.57&2.14 \\
    &    &    &0.50&1.0 &-35.71&0.47&1.57&1.6 &-30.27 &0.44&1.57 \\
    &    &    &     &1.1& -66.52 &0.38&1.35&1.7 &-51.10 &0.36 & 1.30\\\hline\hline

\end{tabular}
\caption{\label{bottom-table3}The continuing of Table
\ref{bottom-table2}.}
\end{table}
\end{center}

\end{document}